\documentclass{emulateapj}

\newcommand{\HeII} {$\rm He\,II$}
\newcommand{\CIII} {$\rm C\,III$}
\newcommand{\CIV}  {$\rm C\,IV$}
\newcommand{\OVI}  {$\rm O\,VI$}

\newcommand{\nO}   {n^{\phantom1}_{\rm O}}
\newcommand{\nH}   {n^{\phantom1}_{\rm H}}

\begin{document}

\title{Anisotropic AGN Outflows and Enrichment of the Intergalactic Medium. 
II. Metallicity} 

\author{Paramita Barai\altaffilmark{1,2,3}, 
Hugo Martel\altaffilmark{2,3}, and 
Jo\"el Germain\altaffilmark{2,3}}

\altaffiltext{1}{Department of Physics and Astronomy,
University of Nevada at Las Vegas, Las Vegas, NV, USA}

\altaffiltext{2}{D\'epartement de physique, de g\'enie physique et d'optique,
Universit\'e Laval, Qu\'ebec, QC, Canada}

\altaffiltext{3}{Centre de Recherche en Astrophysique du Qu\'ebec}

\begin{abstract}

We investigate the large-scale influence of outflows from AGNs 
in enriching the IGM with metals in a cosmological context. 
We combine cosmological simulations of large-scale
structure formation with a detailed model of 
metal enrichment, in which outflows
expand anisotropically along the direction of least resistance,
distributing metals into the IGM. The metals carried by the outflows are
generated by two separate stellar populations: stars located
near the central AGN, and stars located in the greater galaxy. 
Using this algorithm, we performed a series of 5 simulations 
of the propagation of AGN-driven outflows 
in a cosmological volume of size $(128h^{-1}{\rm Mpc})^3$ 
in a $\Lambda$CDM universe, 
and analyze the resulting metal enrichment of the IGM.
We found that the metallicity induced in the IGM is greatly dominated 
by AGNs having bolometric luminosity $L > 10^9 L_{\odot}$, 
sources with $10^8 < L / L_{\odot} < 10^9$ having a negligible contribution. 
Our simulations produced an average IGM metallicity of 
$\left[ {\rm O}/{\rm H} \right] = -5$ at $z = 5.5$, 
which then rises gradually, and remains relatively flat at a value
$\left[ {\rm O}/{\rm H} \right] = -2.8$ between $z = 2$ and $z=0$. 
The ejection of metals from AGN host galaxies by AGN-driven outflows 
is found to enrich the IGM to $> 10 - 20 \%$ 
of the observed values, the number dependent on redshift. 
The enriched IGM volume fractions are small at $z > 3$, 
then rise rapidly to the following values at $z = 0$: 
$6 - 10 \%$ of the volume enriched to $\left[ {\rm O}/{\rm H} \right] > -2.5$, 
$14 - 24 \%$ volume to $\left[ {\rm O}/{\rm H} \right] > -3$, and 
$34 - 45 \%$ volume to $\left[ {\rm O}/{\rm H} \right] > -4$. 
At $z \geq 2$, 
there is a gradient of the induced enrichment, 
the metallicity decreasing with increasing IGM density, 
enriching the underdense IGM to higher metallicities,
a trend more prominent with increasing anisotropy of the outflows. 
This can explain observations of metal-enriched low-density IGM at 
$z\sim3-4$.

\end{abstract}


\section{INTRODUCTION} 
\label{sec-intro} 

Several lines of evidence, both observational and theoretical, 
demonstrate that 
the accretion of matter onto supermassive black holes (SMBHs) 
at the centers of active galaxies, 
and the resulting feedback from them, 
have participated intimately in the formation and evolution of galaxies, and 
strongly influenced their present-day large-scale structures 
\citep[e.g.,][]{salpeter64, lynden-Bell69, rees84, richstone98, ferrarese00, 
kauffmann00, sazonov05, hopkins06, lapi06, malbon07, menci08, shankar09}. 
A large fraction of AGNs are observed to host outflows, 
in a wide variety of forms \citep[e.g.,][]{crenshaw03, chartas09}; 
theoretical studies also indicate triggering of 
massive galactic winds from black holes \citep{king03, monaco05}. 
Searching for AGN feedback, \citet{nesvadba06} observed 
that the outflow from a powerful radio galaxy 
at $z = 2.16$ (likely a forming massive galaxy) 
could remove a significant amount of gas ($\lesssim 50 \%$) 
from a $L>L^{\star}$ galaxy within a few tens to 100 Myr. 
They concluded that AGN winds might have a 
cosmological significance comparable to, or perhaps larger than, 
starburst-driven winds. 
\citet{nesvadba08} found spectroscopic evidence for bipolar outflows in 
three powerful radio galaxies at $z \sim 2 - 3$, with kinetic energies 
equivalent to $0.2\%$ of the rest-mass of the SMBH. 
These outflows possibly indicate a significant phase in the
evolution of the host galaxy. 

AGN outflows can enrich the intergalactic medium (IGM) 
with metals produced inside the host galaxy. It is of great interest to 
study the impact of this
metal enrichment on the evolution of the IGM and the subsequent formation
of galaxies. While the energy deposited by outflows can potentially
inhibit the formation of low-mass galaxies, metals enrichment
can potentially counter-balance this effect by reducing the 
cooling rate of the intergalactic gas.
At present, the full mechanism and operation of AGN feedback and 
metal enrichment on different scales 
is poorly understood, both from theoretical and observational points of view
\citep{mathur09}. 


The metallicity in the 
vicinity of AGNs have been observed to be super-solar, 
and non-evolving over a wide redshift range, indicated by various studies as 
follows. Observational spectroscopy of high-redshift quasars indicate that 
broad-line regions possess gas metallicities several times solar 
\citep[e.g.,][]{hamann92, dietrich03, nagao06a, wang09}, 
reaching as high as $\sim 15 Z_{\odot}$ \citep{baldwin03}. 
\citet{nagao06a} found no metallicity evolution over $2.0 \leq z \leq 4.5$ 
for a given quasar luminosity, 
and estimated the typical metallicity of broad-line region (BLR) 
gas clouds to be $Z \sim 5 Z_{\odot}$. 
\citet{dietrich09} found that the gas metallicity 
of the BLR is $\sim 3 Z_{\odot}$. 
\citet{juarez09} 
inferred that the metallicity of the BLR gas is very high 
(several times solar) even at $z \sim 6$, 
and observed a lack of metallicity evolution among quasars 
within $4 < z < 6.4$ and those at lower-$z$, as found in other studies. 
In addition to the BLRs, narrow absorption line systems 
in QSOs and Seyferts have been observed to have metal abundances 
above the solar value \citep{dOdorico04, groves06}, 
with the gas metallicity in the NLRs 
non-evolving over $1 \leq z \leq 4$ \citep{nagao06b, matsuoka09}. 
Cosmological hydrodynamic simulations \citep{diMatteo04} 
also indicate that quasar hosts have supersolar ($Z / Z_{\odot} \sim 2 - 3$) 
metallicities already at $z \sim 5 - 6$, 
and the rate of evolution of the mean quasar metallicity 
as a function of redshift is generally flat out to $z \sim 4 - 5$. 


The average metallicity of the IGM is found to rise above 
$Z_{\rm IGM} \gtrsim 10^{-3} Z_{\odot}$ by $z \sim 2 - 3$, 
as indicated by several studies, outlined below. 
Observations have detected carbon (\CIV) lines 
in the low neutral hydrogen column density 
Lyman-$\alpha$ forest clouds toward $z \sim 3$ quasars. 
Clouds with $N_{\rm HI} > 10^{15}$ cm$^{-2}$ are found to have a carbon 
abundance of $\sim 10^{-3} - 10^{-2}$ solar \citep{cowie95, songaila96, songaila97}. 
Studying the evolution of intergalactic metals, \citet{songaila01} 
found that the \CIV\ column density distribution function is invariant over 
the redshift range $z = 1.5 - 5.5$, 
and an IGM metallicity of $2 \times 10^{-4} Z_{\odot}$ is already in place at $z = 5$. 
\citet{ryan-weber09}, surveying intergalactic metals between $5.2 < z < 6.2$, 
deduced that the \CIV\ mass density at a mean $\langle z \rangle = 5.76$ 
is smaller by a factor of $\sim 3.5$ compared to the value at $z \leq 4.7$. 
Their study implies an IGM metallicity $Z_{\rm IGM} \gtrsim 10^{-4} Z_{\odot}$. 
Oxygen (\OVI) observations show heavy-element abundances 
in the range $\sim 10^{-3} - 10^{-1.5}$ solar at $z \sim 2 - 2.5$ 
\citep[e.g.,][]{bergeron02, simcoe02, carswell02, simcoe04}. 
Comparing the results of numerical simulations to observations, 
\citet{hellsten97} and \citet{rauch97} showed that 
the observed column density ratios of different ionic species 
at $z \simeq 3$ can be well reproduced 
if a 
ratio $[{\rm C/H}] \simeq -2.5$ is assumed in the models, 
with a scatter of roughly an order of magnitude. 
Such high observed metallicities are explained by 
models involving star formation early on at higher redshifts ($z > 6$), 
including massive star formation \citep{gnedin97}, 
and SN-driven pregalactic outflows 
\citep{madau01, scannapieco02}. 
Some of these earliest metals could have originated in 
Population III stars at very high redshifts 
\citep{carr84, ostriker96, haiman97, wise08}. 
At low redshifts, \OVI\ observations indicate a mean cosmic metallicity of 
$\log(Z_{\rm IGM}/Z_{\odot}) \geq -2.4$ at $z = 0.9$ \citep{burles96}. 
In the local Universe, we have
$[{\rm C/H}] = -1.2$ and $[{\rm O/H}] \gtrsim -2$ \citep{tripp02}, 
metallicities less than $2.5 - 10 \%$ of solar values \citep{shull03}, and a
mean oxygen metallicity of $\sim 0.09 Z_{\odot}$ \citep{danforth05}. 

The IGM enrichment is highly inhomogeneous. 
In cosmological simulations, \citet{cen99} obtained global average 
metallicity values increasing from $1 \%$ of the solar value at $z = 3$ to $20 \%$ at present, 
and noted the very strong dependency on local density, 
with high-density regions having much higher metallicity than low-density regions.
\citet{schaye03} 
found that the carbon abundance in the IGM is spatially highly inhomogeneous 
and is well described by a lognormal distribution for fixed overdensity 
and redshift. Regions of low-metallicity have also been observed: 
\citet{prescott09} discovered a $\approx 45$ kpc Ly-$\alpha$ nebula at 
$z \approx 1.67$, 
exhibiting strong \HeII\ emission and very weak \CIV\ and \CIII] emission, 
which implies low metallicity gas ($Z_{\rm IGM}\lesssim10^{-2}-10^{-3}Z_{\odot}$). 

Several studies, going to as early as \citet{voit96}, 
have showed that winds (starburst, SN-driven, or quasar) from early galaxies 
drive shocks that heat, ionize, and enrich the IGM. 
\citet{tegmark93} modeled SNe-driven winds and 
predicted that most of the IGM is enriched to at least $10 \%$ 
of the current metal content by $z = 5$. 
Adding a prescription for chemical evolution and metal ejection by winds 
in cosmological simulations, \citet{aguirre01} calculated the enrichment 
of the IGM at $z \gtrsim 3$ by galaxies of baryonic mass $\gtrsim 10^{8.5} M_{\odot}$. 
They found that winds of velocity $\gtrsim 200-300$ km s$^{-1}$ can enrich 
the IGM to the mean observed level, 
although many low-density regions would remain metal-free. 
Carrying out SPH simulations, \citet{scannapieco06} showed that the 
observed \CIV\ correlation functions cannot be reproduced by models 
in which the IGM metallicity is constant or a local function of overdensity. 
However, the observed properties are consistent with a model in which 
metals are confined within bubbles with a typical radius $R_s$ 
about sources of mass $ \geq M_s$, 
with best-fitting values of $R_s \approx 2$ comoving Mpc 
and $M_s \approx 10^{12} M_{\odot}$ at $z =  3$. 

On a different note, \citet{gnedin98} showed, using cosmological simulations, 
that most of the metals in the IGM are transported 
from protogalaxies by the merger mechanism, 
and only a small fraction of metals is delivered by SNe-driven winds. 
\citet{ferrara00} showed that SNe explosions and subsequent blowouts 
fail by more than one order of magnitude 
to pollute the whole IGM to the metallicity levels observed, 
requiring some additional physical mechanism(s) to be more efficient pollutants. 
Employing the adiabatic feedback model, 
where stellar feedback naturally drives winds in SPH simulations, 
\citet{shen09} found that IGM metals primarily reside in the 
warm-hot intergalactic medium (WHIM) 
(metallicity of $0.01-0.1~Z_{\odot}$ with a slight decrease at lower-$z$) 
throughout cosmic history. 
They also found that galactic winds most efficiently enrich the IGM 
for intermediate mass galaxies ($10^{10} - 10^{11.5} M_{\odot}$). 

\citet{scannapieco02} performed Monte Carlo cosmological simulations to track 
the time-evolution of SN-driven metal-enriched outflows from early galaxies, 
finding that up to $30 \%$ volume of the IGM 
is enriched to above $10^{-3} Z_{\odot}$ at $z = 3$, 
making the enrichment biased to the areas near the starbursting galaxies 
themselves. The majority of enrichment occurred relatively early 
($5 \leq z \leq 12$), with mass-averaged cosmological metallicity values 
$10^{-3} - 10^{-1.5} Z_{\odot}$. 
Using a hydrodynamic simulations, \citet{thacker02} showed that outflows from 
starbursting dwarf galaxies (with total halo masses $\lesssim 10^{10} M_{\odot}$) 
enrich $\sim 20 \%$ of the simulation volume 
with a mean metallicity of $0.3 \%$ solar at $z = 4$. 
Performing SPH simulations, 
\citet{khalatyan08} found that, 
without AGN feedback, metals are confined to vicinities of galaxies, 
underestimating the observed metallicity of the IGM at overdensities 
$\lesssim 10$, while the AGN feedback model and the efficiency they assumed 
overenrich the underdense IGM. 
\citet{oppenheimer09} performed high-resolution cosmological simulations to study 
the observability and physical properties of five ions 
in absorption between $z = 8$ and $z=5$. 
The volume filling factor of metals they calculated increases during this epoch, 
reaching $\sim 1 \%$ for $Z_{\rm IGM}>10^{-3}Z_{\odot}$ by $z=5$. 
Using hydrodynamic simulations, \citet{tornatore09} found that 
feedback from winds driven by supernovae and BHs leave 
distinct signatures in the chemical and thermal history of the IGM (especially at $z < 3$), 
with BH feedback providing a stronger and more pristine enrichment of the warm-hot IGM. 

While most of the previous work on dispersing metals into the IGM 
has concentrated on winds driven by SNe or starbursts,
some authors have focused on feedback by AGN-driven outflows.
\citet{moll07} examined metal enrichment of the intracluster medium (ICM) 
caused by AGN outflows in galaxy clusters, between $z = 1 - 0$. 
\citet{fabjan10} studied of the effect of AGN feedback on metal enrichment and 
thermal properties of the ICM in hydrodynamical simulations, 
considering feedback from gas accretion onto central SMBHs and that in the `radio mode'. 
Several authors \citep[e.g.,][]{GKW01, barai04, GKWB04, barai07} 
have qualitatively argued that 
the huge lobes of radio galaxies could contribute substantially to spreading 
metals into the IGM, by sweeping out the metal-rich ISM of young galaxies 
which they encounter while expanding. 
In a recent study, 
direct observational evidence for outflow of metal-enriched gas 
driven by a radio AGN were found by \citet{kirkpatrick09}. 
Using Chandra observations of the Hydra A galaxy cluster, they showed that the 
metallicity of the ICM is enhanced by up to 0.2 dex along the radio jets and lobes 
(enhancements extending over a distance of $20 - 120$ kpc from the central galaxy) 
compared to the metallicity of the undisturbed gas. 
They estimated that $10 - 30 \%$ of the iron mass 
within the central galaxy has been transported out. 

Other cosmological studies on quasar/radio galaxy outflows 
\citep{FL01, so04, lg05, barai08} 
have considered outflows expanding with a spherical geometry. 
However, in realistic cosmological scenarios, where the density distribution 
shows significant structures in the form of filaments, pancakes, etc., 
outflows are expected to expand anisotropically on large scales. 
In a previous paper (\citealt{GBM}, hereafter Paper~I), 
we designed a semi-analytical model for AGN outflows expanding with 
an anisotropic geometry, following the path of least resistance, 
which we implemented into a numerical algorithm for cosmological simulations. 
We used this algorithm to study the distribution and enriched volume fraction
of metals produced by a cosmological population of AGNs,
in a $\Lambda$CDM universe, over the age of the Universe. These
results are presented in Paper~I.
In the second part of this project, 
we have modified the algorithm used in Paper~I by adding a prescription 
for the
amount and composition of the metals transported by AGN outflows.
Using this modified algorithm, we now complete the study of Paper~I by
calculating the metallicity and chemical composition of the IGM.

This paper is organized as follows. 
In \S\ref{sec-numerical}, we describe the numerical methodology for 
incorporating a metal enrichment technique into our anisotropic outflow model. 
\S\ref{sec-cosmol-sims} describes the cosmological $N$-body simulations 
we performed. 
The results are presented and discussed in \S\ref{sec-results}. 
We summarize and present our conclusions in \S\ref{sec-conclusion}. 

\section{THE NUMERICAL METHOD} 
\label{sec-numerical} 




\subsection{The Basic Algorithm}
\label{sec-BasicAlgo} 

Our basic algorithm is described in details in Paper~I. We use a
Particle-Mesh (PM) code to simulate the growth of large-scale structure
in the universe, in a cubic volume with periodic boundary conditions
expanding with Hubble flow. At each time step, we filter the density
distribution to identify the density peaks (local maxima), where
AGNs will be located. We assume that each AGN has an active lifetime
$t^{\phantom1}_{\rm AGN}=10^8{\rm yr}$, and that AGN luminosities range from
$10^8L_\odot$ to $10^{14}L_\odot$ with the luminosity distribution function
of \citet{hopkins07}. We then determine the number of AGNs
and their birth redshifts in order to reproduce the redshift-dependent
AGN distribution function $N(L,z)$. As the simulation proceeds, we
locate, at each time step, the appropriate number of newly-born AGNs at 
randomly-selected density peaks. Using scaling relations between
the mass $M_{\rm bulge}$ of the bulge and the mass  $M_{\rm BH}$
of the central black hole, and between $M_{\rm BH}$ and the AGN luminosity,
we estimate the baryonic mass $M_b$ of the host galaxy, and
its total mass $M_{\rm gal}=M_b\Omega_0/\Omega_{b,0}$,
where $\Omega_{b,0}$ and
$\Omega_0$ are the baryon and total density parameters, respectively.
We assume that a fraction 
$f_{\rm outflow}=0.6$ of these AGNs produce outflows
\citep{gb08}. 
Our outflow model is based on the outflow evolution equations
given in \citet{tegmark93}, combined with the anisotropic outflow model of
\citet{pmg07}. 
In this model outflows form two bipolar cones of
opening angle~$\alpha$, expanding along the direction of least
resistance around the density peak. We treat the opening angle as a 
free parameter. For details, we refer the reader to Paper~I. 

We note that using an analytical outflow model necessarily requires 
that we make some simplifying assumptions. 
The model of \citet{tegmark93} 
assumes that the external medium has an uniform density. 
Following it, our model of outflow evolution approximates 
that the density of the environment through which the AGN outflows propagate 
is equal to the mean baryon density of the universe, at the corresponding redshift. 
In a general scenario, the density of the external medium would vary 
both with distance and direction. 
However, 
the directional dependence of the density is taken into account in our model, 
by using anisotropic outflows that travel along the direction of least resistance. 
In that sense, our model constitutes an improvement over that of \citet{tegmark93}. 
The distance dependence of the density is taken into account in the 
outflow expansion model used by \citet{lg05}, 
which is also an improvement over \citet{tegmark93}. 
Our model does not take into account of this distance dependence. 
Rather, the focus of our study is on the directional dependence of the density field, 
and the resulting anisotropic outflows. 
Expanding along the path of least resistance, 
our model outflows travel in regions that are first overdense, 
then underdense, hence the effect of the density variations partly cancel out.


This basic algorithm enabled us to identify the regions in space
that were metal-enriched by outflows, and to calculate the enriched volume
fraction (Paper~I), but did not consider the actual amount of metals
transported by the outflows. We have therefore modified this algorithm
to take into account the mass and chemical composition of the metals
contained in each outflow, in order to calculate the metal content and
metallicity of the enriched IGM. The modifications to the basic algorithm
are described in \S\S2.2--2.4 below.

\subsection{Mass of Metals in Galaxy} 
\label{sec-MetalMassGal} 

Metals are generated within a galaxy by stars undergoing SNe explosions. 
The AGN outflows act to spread the produced metals to the broader IGM. 
We distinguish two regions of stellar populations as the sources of metal 
generation: 
(1) stars near the AGN (very center of galaxy), and 
(2) stars located away from central AGN within the rest of the galaxy. 

The observational diagnosis of the metallicities is different in the two 
regions, which makes it difficult to disentangle them at high redshifts. 
Observational studies of AGN/quasar metal abundances measure metallicities 
of the 
broad line and narrow line gas clouds near the AGN, which have a high metal 
content (region 1). 
Studies of galactic metallicity measure the abundances in the larger 
galaxy ISM (region 2), which are generally lower than in region 1. 
Previous simulations by \citet{moll07} assumed that only the central 
metals (region 1) are driven out to the IGM by AGN outflows. 
They neglected the fact that AGN outflows can entrain some enriched ISM 
of the larger galaxy and spread it to the IGM. 
In this work, we take into account both sources of metals in regions~1 
and~2. 

Quasar abundances studies indicate that the mass of gas enriched in metals 
near the central engine is at least equal to the mass of the central black 
hole, $M_{\rm gas} \geq M_{\rm BH}$ \citep{hamann07}. 
In this work, we adopt the lower limit ($M_{\rm gas}=M_{\rm BH}$) 
as the mass of metal-enriched gas 
near the AGN (region 1) which is carried by the outflow.
In the absence of well-constrained values and for simplicity, 
we assume a central metallicity of 
$Z_1 = 5 Z_{\odot}$ for all the AGN in our simulations. 
Such an assumption is supported by spectroscopic observations of 
quasars (\S\ref{sec-intro}), 
which have detected several times solar metallicity of broad-line regions, 
with no metallicity evolution over a broad redshift range. 
\citet{moll07} also used a value of $5 Z_{\odot}$ as their outflow 
metallicity. 
Hence, the mass of metals in the central region  
which is carried away by an outflow and dispersed into the IGM is 
given by $M_{Z,1}=Z_1 M_{\rm BH}=5Z_\odot M_{\rm BH}$. 
We take the mass fraction of heavy elements in the Sun 
as $Z_{\odot} = 0.02$ \citep{cox00}, 
to be consistent with the value used in our analysis (detailed in Appendix~A) 
for computing the metal mass fraction. 
Such values are based on older estimates of the Solar metal abundance 
\citep{anders89}, 
and have also been used by other studies with which we compare our results. 

For region 2, 
a metallicity value is ascribed to each AGN host galaxy according to its mass.
We adopt the estimation of redshift-dependent mass-metallicity relation 
by \citet{maiolino08}, where the following formulation is used, 
\begin{equation} 
\label{eq-metal} 
12 + \log(\nO/\nH)=-0.0864 (\log M_{\star} - \log M_0)^2 + K_0\,,
\end{equation} 
where the metallicity is expressed as the 
logarithm of oxygen over hydrogen abundance (number density ratio), 
$\log (\nO/\nH)$, 
and $M_{\star}$ is the galaxy stellar mass.\footnote{\citet{maiolino08}
use the notation ${\rm O/H}$ for $\nO/\nH$.} 
$M_0$ and $K_0$ are fitting parameters whose values are 
determined at each redshift to obtain best fit to the observations. 
Table 5 of \citet{maiolino08} gives the parameter values 
over the redshift range $z = 0.07 - 3.5$. 
For a given redshift $z$, we obtain the parameters by interpolating 
between two consecutive lines in that table. 
For redshifts $z < 0.07$ and $z > 3.5$, 
we simply use the values at $z=0.07$ and $z=3.5$, respectively. 
The stellar mass and ISM mass of the galaxy are
$M_{\star} = f_{\star} M_b$ and $M_{\rm ISM} = (1-f_{\star}) M_b$, 
respectively,
where $f_{\star}$ is the star formation efficiency.
We use the value $f_{\star} = 0.1$, as in \citet{pmg07}.

Using the value of $\nO/\nH$
obtained from equation~(\ref{eq-metal}),
we then calculate the mass ratio of metals 
using
\begin{equation} 
\label{eq-ZG} 
Z_G=\frac{\mu^{\phantom1}_{Z,{\rm O}}}
{(\nO/\nH)^{-1}
\mu^{\phantom1}_{\rm pr,H} + \mu^{\phantom1}_{Z,{\rm O}}}\,,
\end{equation} 

\noindent where
\begin{equation} 
\label{mu-equiv}
\mu^{\phantom1}_{Z,{\rm O}}\equiv\sum_{Z>2}
\frac{n^{\phantom1}_Z}{\nO}\,\mu^{\phantom1}_Z\,,\qquad
\mu^{\phantom1}_{\rm pr,H}\equiv\mu^{\phantom1}_{\rm H}
+\frac{n^{\phantom1}_{\rm He}}{\nH}
\,\mu^{\phantom1}_{\rm He}\,, 
\end{equation}

\noindent and the sum in equation~(\ref{mu-equiv}) is over all metals
(all elements except hydrogen and helium).
The derivation of these relations is presented in Appendix~A.
 
For each AGN, the metallicity $Z_G$ of its host galaxy is computed using 
equations~(\ref{eq-metal})--(\ref{mu-equiv}).
$Z_G$ is then
multiplied by the gas mass in the galaxy ISM ($M_{\rm ISM}$) 
to get the mass of metals $M_{G, en}$ in the ISM. 
Only these metals can be carried by an outflow, since
the metals locked up inside stars cannot escape. We assume that
a fixed fraction $f_{\rm esc}$ of the metals in the ISM will be carried by 
the outflow (if there is an outflow), and we 
use the value of $f_{\rm esc} = 0.5$, as in \citet{pmg07}. 
The mass of metals carried away by an outflow 
from the non-central regions of its host galaxy 
is then given by $M_{Z, 2} = f_{\rm esc} M_{G,{\rm en}} 
= f_{\rm esc} Z_G (1-f_{\star}) M_b$. 
The total mass of metals carried by the outflow is therefore
\begin{equation}
M_{Z,{\rm out}}=5Z_\odot M_{\rm BH}+f_{\rm esc} Z_G (1-f_{\star}) M_b .
\label{mzout}
\end{equation}



\subsection{Distribution of Metals from Galaxy to IGM} 
\label{sec-MetalTransfer} 

We simulate the
enrichment of the IGM by 
distributing the metals transported by the outflows to the 
particles (of the PM code) intercepted by them. 
The exact mechanisms by which 
galactic metals are distributed by outflows to the IGM are not well-known, 
so we have to make some simplifying approximations. 
We assume that the metals carried by the outflow escape the
galaxy at a constant rate during the active-AGN life, 
starting from the time $t_{\rm bir}$ when the outflow is born 
until the time $t_{\rm off}$ when the central engine turns off. 
The total mass of metals carried from the central AGN and the greater galaxy, 
$M_{\rm Z, out} = M_{Z, 1} + M_{Z, 2}$, 
is considered to be deposited uniformly into the IGM gas 
overlapped by the outflow.
We implement this into the algorithm as follows.

1) Each outflow will deposit its total metal content 
$M_{\rm Z, out}$ into the diffuse IGM, 
over the time the AGN is active (i.e., from birth to switch-off epoch).
The time during which the outflow expands in the active-AGN phase is
$t^{\phantom1}_{\rm AGN}=t_{\rm off}-t_{\rm bir}=10^8~{\rm yr}$. 
The metals carried by the outflow are considered to be deposited 
into the IGM at an uniform rate in time during the active lifetime.
Hence, during the active-AGN phase, an amount of metals 
$M_{\rm Z, out}\Delta t/t^{\phantom1}_{\rm AGN}$ is added to the outflow 
during a timestep $\Delta t$. 
If at that time the outflow intercepts $N_p$ particles,
the mass of metals added to each intercepted particle
is then $M_{\rm Z, out} \Delta t / (t^{\phantom1}_{\rm AGN} N_p)$. 
This insures that the total mass of 
metals deposited in the IGM during the active-AGN life is equal to 
$M_{\rm Z, out}$.

2) Simply following such a prescription (point 1) leads to the following 
issue. 
When a metal-enriched outflow permeates ambient matter, 
the parts of IGM being permeated first gain more metals than 
the parts permeated later during the expansion of the outflow. 
In our simulations, this adds more metals to the particles near the AGN 
(which are overlapped by the expanding outflow starting from a time close to 
$t_{\rm bir}$), than the particles located far away 
(which are intercepted later at a time close to $t_{\rm off}$), as the outflow 
expands. 
This might indeed be the case in reality 
[the model of \citep{tegmark93} assumes a linear velocity field 
inside the outflow, which implies no mixing]. 
However, turbulent motion inside the outflow could redistribute the metals. 

In this paper, we assume that the
distribution of metals in the outflow is uniform at all times. To
implement this 
we redistribute the metals within the outflow volume, 
by performing an averaging of the metal mass contained by the 
relevant particles. At each timestep,
we identify all the $N_p$ particles intercepted by the outflow, 
add up the metals (as explained in point~1 above),
and redistribute these metals evenly among the $N_p$ particles.
Such a procedure smooths out the metal distribution in the IGM 
volume overlapped by the outflow, such that at every timestep, all particles
inside the outflow have the same metal content, which steadily
increases during the active AGN life as more metals are added.

3) After the active-AGN phase is completed (at time $t=t_{\rm off}$), 
all the metals carried by the outflow have been deposited into the IGM.
At that point, the outflow enters a post-AGN dormant phase
(see \S2.5 of Paper~I). The internal pressure of the outflow exceeds
the external pressure of the IGM, driving the expansion further. During
that phase, more IGM gas is being swept by the outflow (in practice, this
means more particles are overlapped by the outflow), but the metal
content of the outflow no longer increases.
To implement this 
we follow the same averaging technique 
(detailed in point 2 above) for this redistribution of metals. 
The particles newly overlapped (after $t_{\rm off}$) 
gain a fraction of metals from the particles that were overlapped
earlier (between $t_{\rm bir}$ and $t_{\rm off}$), 
such that the total metal mass in the outflow is conserved. Eventually,
the outflow reaches pressure equilibrium with the external IGM. 
At that point, the outflow passively follows the Hubble flow, 
and no new IGM gas is being swept.

We need to be careful in situations when outflows overlap. If a particle
is hit by several outflows, it should have a higher metal content than 
particles hit by only one outflow. However, the averaging prescription
described above would tend to erase this, by giving to all
particles inside a same outflow the same metal content.
The proper way to deal with this situation would be to keep track
of the amount of metals each particle has received from each outflow.
For instance, suppose that a particle has been enriched by two outflows
$A$ and $B$, which have deposited an amount of metals $M_{Z,A}$ and
$M_{Z,B}$ onto that particle, respectively. When averaging the metal
content of particles inside outflow $A$, only the mass $M_{Z,A}$ would
be involved in the averaging. Similarly, when averaging the metal
content of particles inside outflow $B$, only the mass $M_{Z,B}$ would
be involved in the averaging. This would conserve the total mass of metals,
while insuring that particles hit by several outflows have a higher
metal content. 

In practice, this is extremely difficult to implement in the algorithm.
We have considered several approaches, and all of them
require either the memory or the
CPU time to scale like the product 
$\hbox{(number of particles)}\times\hbox{(number of outflows)}$,
that is, $256^3\times1.5\times10^6$.
To deal with this situation, we make the approximation that
$M_{Z,A}=M_{Z,B}$. More generally,
if a particle $P$ overlaps with $N$ outflows, we assume that each outflow
contributed a fraction $1/N$ of the metals contained in that particle.
Then, when applying the averaging technique to one of these outflows,
only a fraction $1/N$ of the metals contained in particle $P$ is
included in the averaging. The result is that, within
each outflow, the particles hit by that outflow only will have the same metal
content, while particles hit by $N$ outflows will contain
roughly $N$ times more metals. This might result in some local fluctuations
when particles are overlapped by outflows having very different
metal content, but with so many outflows in our computational volume, the
overall effect of this approximation should be small. 
Also, a more luminous outflow (higher $M_{\rm BH}$ and $M_b$) would carry more metals (high $M_{\rm Z, out}$), 
but at the same time would grow larger, 
intercepting more particles (larger $N_p$) within its volume than a less luminous outflow. 
So the amount of metals deposited to a particle 
$M_{\rm Z, out} \Delta t / (t^{\phantom1}_{\rm AGN} N_p)$ 
remains comparable for outflows with different luminosities. 
Notice that the 
total mass of metals is strictly conserved in this procedure.




The methodology described in \S\S\ref{sec-MetalMassGal} 
and \ref{sec-MetalTransfer}
is implemented in the PM code, hence the propagation of outflows and the 
enrichment of the IGM is simulated along with the growth of large-scale 
structure.
In the next section (\S\ref{sec-computeMetal}), we describe calculations
that use the dumps produced by the simulations
(which contain the positions, velocities, masses, and metal content of
the PM particles). These calculations
are performed as post-processing after the simulations are completed. 

\subsection{Computation of the Metallicity of the IGM} 
\label{sec-computeMetal} 

The outflowing matter 
consists of a mixture of primordial gas (H, He) 
and metals carried from the host galaxy. 
The mass of H and He transferred by an outflow 
to the overlapped IGM volume is assumed to be negligible 
compared to the mass of H and He already present in the IGM. 
Only the mass of metals deposited into the IGM is considered to 
be significant and counted. 
Any IGM volume is considered to have no metals 
before it has been encompassed by an outflow. Hence, in the initial
conditions, the metal content of each PM particle is set to zero.

In order to compute the metallicity of the IGM at a given time (or redshift),
we take the corresponding dump, which contains the particle positions, masses,
and metal content. 
We then divide the computational volume into 
$N_{\rm ff}^3 = 256^3$ cubic cells 
and we compute the matter density $\rho_{\rm cell}$ and metal density
$\rho_{\rm metal}$ in the center of each cell, 
using a 
smoothing method, as described in Paper~I. 
We then multiply these numbers by
the cell volume to get the total mass $M_{\rm cell}$ and total metal 
mass $M_Z$ in each cell.
Assuming a universal ratio of baryons to dark matter, 
the mass of baryonic gas (comprising of H, He, and metals) in the cell is 
$(\Omega_{b,0} / \Omega_0) M_{\rm cell}$. 

Once the AGN outflows are born and start propagating into the IGM, 
the IGM is enriched by metals carried by the outflows. 
If $M_{Z}$ is the mass of metals added to a grid cell (from outflows
 hitting it), the metal abundance ratio, by mass, of the cell is 
\begin{eqnarray} 
\label{eq-Zcell} 
Z_{\rm cell} & = & 
\frac{M_{Z}} {  M_{\rm H} + M_{\rm He} + M_{Z} } \approx \frac{ M_{Z} } { M_{\rm H} + M_{\rm He}  } \nonumber \\ 
& \approx & \frac{ M_{Z} } { \left( M_{\rm H} + M_{\rm He} \right)_{\rm initial} } = \frac{ M_{Z} }{(\Omega_{b,0}/\Omega_0)M_{\rm cell} }. 
\end{eqnarray} 
Here it has been assumed that the mass of metals imparted to a cell 
is negligible compared to the mass of H and He already there 
($M_{Z}\ll M_{\rm H} + M_{\rm He}$), 
and that the total mass of H and He in the cell remains the same 
before and after outflow impact, the added mass being negligible. 

Using equation~(\ref{eq-ZG}), we can calculate the
number density ratio of O over H,
\begin{equation}
\label{eq-ObyHcell} 
\left(\frac{\nO}{\nH}\right)_{\rm cell} = 
\frac{\mu^{\phantom1}_{\rm pr,H}}{\mu^{\phantom1}_{Z,{\rm O}}} 
\left( \frac{Z_{\rm cell}}{1-Z_{\rm cell}} \right). 
\end{equation} 
This gives us the oxygen abundance ratio in each grid cell of the 
computational volume. 

We finally express the metal content in terms of the metallicity, 
which is the logarithmic ratio of the metal abundance of a cell over the 
corresponding solar value: 
$[{\rm O}/{\rm H}]_{\rm cell} = 
\log_{10}(\nO/\nH)_{\rm cell} - 
\log_{10}(\nO/\nH)_{\odot}$. 
We compute the metallicity using oxygen, as it is the most abundant element 
(after hydrogen and helium) in the universe. 
The metallicity for any other element can be similarly computed, 
assuming some abundance ratio. 

\section{THE COSMOLOGICAL SIMULATIONS}
\label{sec-cosmol-sims} 


\begin{deluxetable}{lrcc} 
\tablecaption{Parameters of the Simulations} 
\tablewidth{0pt} 
\tablehead{ 
\colhead{Run} & 
\colhead{$\alpha$ ($^{\circ}$)} & 
\colhead{$L_{\min} / L_{\odot}$} & 
\colhead{$L_{\max} / L_{\odot}$}  
} 
\startdata 
A  & $180$ & $10^8$ & $10^{14}$ \cr 
B  & $120$ & $10^8$ & $10^{14}$ \cr 
C  & $60$  & $10^8$ & $10^{14}$ \cr 
C1 & $60$  & $10^9$ & $10^{14}$ \cr 
C2 & $60$  & $10^8$ & $10^9$    \cr 
\enddata 
\label{TabRuns} 
\end{deluxetable} 

As in Paper I, we simulate the growth of large-scale structure and the
propagation of outflows in a cubic cosmological
volume of comoving size $L_{\rm box} = 128\,h^{-1}$ Mpc $= 182.6\,{\rm Mpc}$
with periodic boundary conditions,
using a Particle-Mesh (PM) algorithm \citep{he88}, with $256^3$ equal mass
particles and a $512^3$ grid. This corresponds to a particle mass
$m_{\rm part}=1.38\times10^{10} M_\odot$,
and a grid spacing $\Delta=0.357\,{\rm Mpc}$.

We consider a $\Lambda$CDM model with a present
baryon density parameter $\Omega_{b,0}=0.0462$,
total matter (baryons + dark matter) density parameter $\Omega_0=0.279$,
cosmological constant $\Omega_{\Lambda,0}=0.721$,
Hubble constant $H_0=70.1\rm\,km\,s^{-1}Mpc^{-1}$ ($h=0.701$),
primordial tilt $n_s=0.960$, and CMB temperature $T_{\rm CMB}=2.725$,
consistent with the results of {\sl WMAP5}
combined with the data from baryonic acoustic oscillations and supernova
studies\footnote{http://lambda.gsfc.nasa.gov/product/map/dr3/parameters\_summary
.cfm}
\citep{hinshaw08}. We generate initial conditions at
redshift $z=24$, and evolve the cosmological volume up
to a final redshift $z=0$. The simulations presented in this paper
use the same initial conditions as simulations~A, B, and~C of Paper~I.

We performed a series of 5 simulations by varying some parameters of our AGN
evolution model.
Table~\ref{TabRuns} summarizes the characteristics of each run.
The first column gives the letter identifying the run. 
In columns 2, 3, and 4, we list the opening angle of the outflows, and
the minimum and maximum bolometric luminosities used for generating 
the AGN population from the QLF (\S2.2, Paper~I), respectively. 
These five simulations were all presented (among others) in Paper~I.
We redid them with our modified algorithm
which incorporates 
the explicit metal enrichment prescription described in 
\S\ref{sec-numerical}. 

In addition, we performed an extra simulation run investigating a different prescription 
for the initial location of the AGN population, whose results are given in \S\ref{sec-InitLoc}.

\section{RESULTS AND DISCUSSION}
\label{sec-results} 

The results are presented and discussed in the following subsections. 
We compute various statistics of the metallicity of oxygen 
(following the prescriptions in \S\ref{sec-computeMetal}) 
to quantify the resulting IGM metal enrichment in our simulations. 

\subsection{The Final Radius of the Outflows}

\begin{figure} 
\centering
\includegraphics[width = \linewidth]{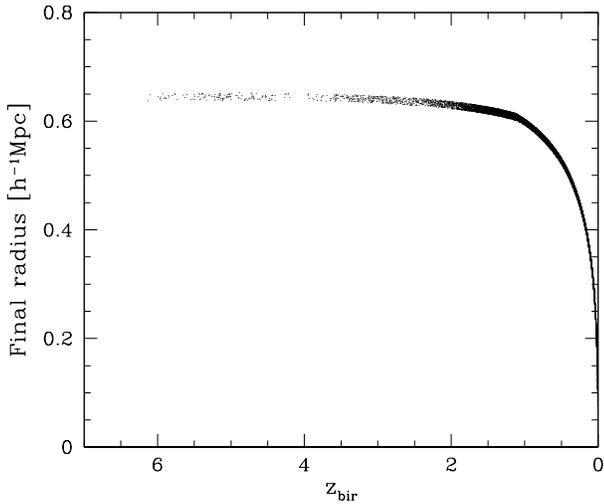} 
\caption{ 
Final radius $R$ of the
outflows vs. birth redshift $z_{\rm bir}$, for galaxies in the mass range
$M_{\rm gal}=[1.0,1.1]\times10^7M_\odot$.
Each dot represents one outflow. 
}
\label{rz}
\end{figure} 

The final radius $R$ of an outflow depends on the mass $M_{\rm gal}$
and birth redshift $z_{\rm bir}$ of the galaxy producing that outflow. 
Examining these dependencies will help us to understand the results 
presented in the following sections.

We consider Run~C. Figure~\ref{rz} shows the final radius of the
outflows vs. birth redshift, for galaxies in the mass range
$M_{\rm gal}=[1.0,1.1]\times10^7M_\odot$. The thickness of the curve
is caused by the use of a finite mass interval. The radius is nearly
independent of redshift for outflows with $z_{\rm bir}>2$. This may seem
surprizing at first. Outflows that form earlier are facing a much 
stronger external IGM pressure, and thus reach pressure equilibrium much 
sooner. However, once they have reached that equilibrium, they do not stop.
Instead, they expand with Hubble flow. For outflows that
reach equilibrium, the final physical radius is equal to the comoving radius
at equilibrium. When outflows reach equilibrium early, their
physical radius at that time is small, but their comoving radius is large.

There is a slight break in the slope at redshift $z=1.1$, followed by a rapid 
decline down to $R=0$ at $z=0$. These outflows do not reach equilibrium,
and are still in the active-AGN phase or
the post-AGN phase by the present time. The closer to the present
the galaxies are born, the less time outflows have to propagate.

\begin{figure} 
\centering
\includegraphics[width = \linewidth]{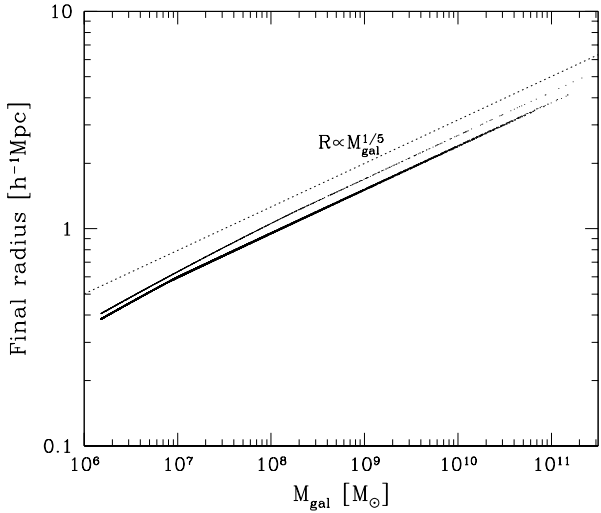}
\caption{ 
Final radius $R$ of the
outflows vs. galaxy mass $M_{\rm gal}$, for galaxies born in
the redshift range
$z_{\rm bir}=[3.1,3.0]$ (middle curve) and 
$z_{\rm bir}=[1.1,1.0]$ (bottom curve).
Each dot in these curves represents one outflow.
The top line shows a power-law approximation to these curves.
}
\label{rm}
\end{figure} 

Figure~\ref{rm} shows the final radius of the outflows vs. galaxy mass,
for galaxies born in the redshift ranges $z_{\rm bir}=[3.1,3.0]$ and 
$z_{\rm bir}=[1.1,1.0]$. More massive galaxies produce outflows that
travel larger distances. However, this dependence is very weak.
We find that $R$ scales approximately like $M_{\rm gal}^{1/5}$.
This is a result of competing effects. 
Following equation~(9) of Paper~I, the thermal pressure generated 
by the AGN luminosity is what drives the expansion of the outflow. 
However, the outflow is decelerated by the gravitational attraction 
of the galaxy hosting the AGN. Both effects increase with galaxy mass, 
resulting in competition.

\subsection{Total Masses in Simulation Volume} 
\label{sec-totMass} 

\begin{figure} 
\centering
\includegraphics[width = \linewidth]{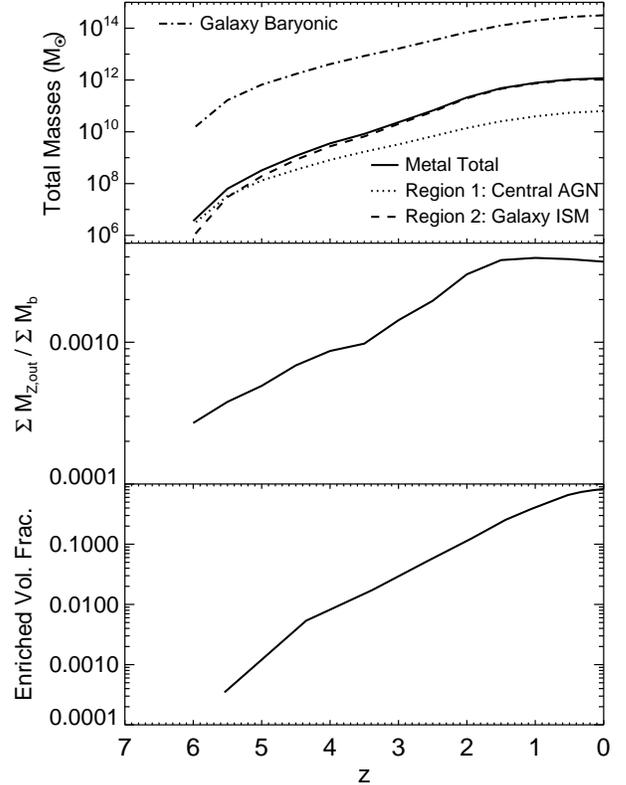}
\caption{ 
Quantities in the whole ($182.6$ Mpc)$^3$ simulation volume 
as a function of redshift. 
The {\it top} panel shows the total baryonic mass of all host galaxies
($\sum M_b$, {\it dot-dashed} upper curve),
and the total mass of metals transported by outflows 
from all galaxies ($\sum M_{\rm Z, out}$, {\it solid curve}). 
The total metal mass is a summation of 
metals carried from the central regions ($\sum M_{Z,1}$, {\it dotted curve}), 
and metals carried from the non-central regions (the greater ISM)
($\sum M_{Z,2}$, {\it dashed curve}).
The {\it middle} panel gives the ratio of metal mass over galaxy mass. 
The upper two panels are the same for runs A, B and C. 
The {\it bottom} 
panel shows the fractional volume 
of the simulation box enriched by the outflows in run C ($\alpha = 60^{\circ}$). 
}
\label{fig-totMass}
\end{figure} 

\begin{figure} 
\centering
\includegraphics[width = \linewidth]{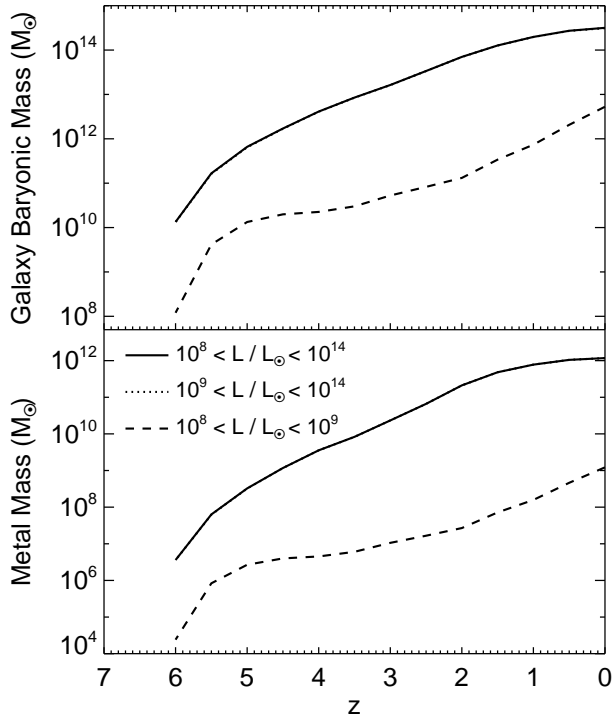}
\caption{ 
Total masses in the whole ($182.6$ Mpc)$^3$ simulation volume 
as a function of redshift, for AGNs with different bolometric luminosity limits: 
the whole population, $10^8 < L / L_{\odot} < 10^{14}$ ({\it solid}), 
only the high-luminosity sources, $10^9 < L / L_{\odot} < 10^{14}$ ({\it dotted}), 
only the low-luminosity sources, $10^8 < L / L_{\odot} < 10^9$ ({\it dashed}). 
The {\it top} panel shows the total baryonic mass of the host galaxies 
($\sum M_b$) within the relevant luminosity range. 
The {\it bottom} panel shows the total mass of metals transported by outflows 
from those galaxies ($\sum M_{\rm Z, out}$). 
}
\label{fig-highLlowLmass}
\end{figure} 


Figure~\ref{fig-totMass} shows the 
redshift evolution of the total baryonic mass contained in galaxies 
hosting AGNs, the total mass of metals carried by outflows, 
and the enriched volume fraction, 
inside the whole ($182.6$ Mpc)$^3$ simulation volume. 
The masses were computed by summing over AGNs with birth 
redshifts $z_{\rm bir}\geq z$.
The top panel shows various total masses: 
the baryonic mass of all host galaxies, $\sum M_b$, 
and the mass of metals transported by outflows from all galaxies,
$\sum M_{\rm Z, out}$,
which includes
metals carried from the central AGN, $\sum M_{Z,1}$, 
and metals carried from the non-central ISM, $\sum M_{Z,2}$
[see eq.~(\ref{mzout})].
The middle panel depicts the ratio of total metal mass over galaxy mass.
Note that these total masses and their ratio do not depend on the outflow 
opening angle $\alpha$. 
Thus the upper two panels are the same for runs A, B, and C. 
The bottom panel gives the fractional volume 
of the simulation box enriched by the outflows in run C 
($\alpha = 60^{\circ}$). 
This enriched volume fraction is the same as the solid curve 
in Figure~4 of Paper~I, but is plotted here on a logarithmic scale. 

The total (dark matter + baryon) mass inside our simulation volume 
is $M_{\rm box} = 2.32 \times 10^{17} M_{\odot}$. 
Scaling by the cosmological baryon to total matter density ratio, 
this corresponds to a total baryonic mass in the box: 
$M_{b,{\rm box}} = M_{\rm box} (\Omega_{b,0} / \Omega_0) 
= 3.84 \times 10^{16} M_{\odot}$. 
We can read out the values of masses from Figure~\ref{fig-totMass}. 
At the present epoch: 
the total baryonic mass in galaxies, 
$\sum M_b (z = 0) = 3.18 \times 10^{14} M_{\odot}$, 
and the total mass of metals available to be distributed to the IGM, 
$\sum M_{\rm Z, out} (z = 0) = 1.18 \times 10^{12} M_{\odot}$. 
The ratio of total baryonic mass of galaxies 
over that of the simulation box is $0.00828$. 
The total metal mass is a fraction $0.00370$ of the baryonic mass in galaxies, 
and $3.07 \times 10^{-5}$ of baryonic mass in the simulation box. Notice that
only galaxies that host AGNs are included in our calculations. Galaxies comprise
of order 5\% of the total mass in the universe, hence galaxies hosting AGN
represent a fraction $0.00828/0.05=17\%$ of all galaxies by mass. 
This is a direct consequence of the QLF we adopted for generating the AGN sources 
to populate our simulation volume and our assumption of the BH-galaxy mass 
relation (\S\ref{sec-BasicAlgo}). 

To study the relative contributions of low- and high-luminosity AGNs on
metal enrichment, we divided the AGN population in two separate luminosity ranges,
$10^8 < L / L_{\odot} < 10^9$ (low-luminosity or, faint AGNs) and 
$10^9 < L / L_{\odot} < 10^{14}$ (high-luminosity or, bright AGNs). 
Figure~\ref{fig-highLlowLmass} shows the evolution of the baryonic mass in galaxies 
and the metal mass separately for low- and high-luminosity AGNs.
We note that the total masses for 
the whole population ($10^8 < L / L_{\odot} < 10^{14}$) 
is completely dominated by the high-luminosity sources. 
Hence the solid and dotted curves in Figure~\ref{fig-highLlowLmass} almost coincide, 
and are not distinguishable in this plot. 
The low-luminosity AGNs ($10^8 < L / L_{\odot} < 10^9$) 
have a very small contribution to the total masses, 
at $z = 0$ they comprise of $1.7 \%$ of the total galaxy baryonic mass 
and $0.1 \%$ of the total metal mass. 
Summed over the entire population of fainter sources, 
the total mass of galaxies is: 
$\sum_{\rm faint} (M_{\rm gal}) = 5.29 \times 10^{12} M_{\odot}$, 
and the total mass of metals available to be distributed to the IGM: 
$\sum_{\rm faint} (M_{\rm Z, out}) = 1.22 \times 10^{9} M_{\odot}$. 
The corresponding numbers for the brighter sources are: 
$\sum_{\rm bright} (M_{\rm gal}) = 3.12 \times 10^{14} M_{\odot}$, 
$\sum_{\rm bright} (M_{\rm Z, out}) = 1.175 \times 10^{12} M_{\odot}$. 
This point is relevant to explain the results about low- and high-luminosity AGNs 
we present in \S\ref{sec-avgMetal} and \S\ref{sec-enrichVol} below. 


\begin{deluxetable*}{crccccc} 
\tablecaption{Galaxy and metal mass for AGNs with different 
bolometric luminosities} 
\tablewidth{0pt} 
\tablehead{ 
\colhead{$(L_{\min}-L_{\max})\>[L_{\odot}]$} & 
\colhead{$N_{\rm AGN}$} & 
\colhead{$(\sum M_{\rm b})\>[M_{\odot}]$} & 
\colhead{$(\sum (M_{\rm Z,out})\>[M_{\odot}]$} &
\colhead{$\%N$} &
\colhead{$\%M_{\rm b}$} &
\colhead{$\%M_{\rm Z,out}$}
} 
\startdata 
$10^{8}-10^{9}$   & 1178571 & $5.29 \times 10^{12}$ & $1.22 \times 10^{9}$ 
& 76.8 &  1.7 &  0.1 \cr 
$10^{9}-10^{10}$  &  264676 & $1.22 \times 10^{13}$ & $4.86 \times 10^{9}$
& 17.2 &  3.9 &  0.4 \cr 
$10^{10}-10^{11}$ &   68655 & $3.24 \times 10^{13}$ & $3.45 \times 10^{10}$
&  4.5 & 10.2 &  2.9 \cr
$10^{11}-10^{12}$ &   19372 & $9.10 \times 10^{13}$ & $2.49 \times 10^{11}$ 
&  1.3 & 28.7 & 21.2 \cr 
$10^{12}-10^{13}$ &    3978 & $1.54 \times 10^{14}$ & $7.44 \times 10^{11}$ 
&  0.3 & 48.5 & 63.2 \cr 
$10^{13}-10^{14}$ &     110 & $2.26 \times 10^{13}$ & $1.43 \times 10^{11}$ 
&  0.0 &  7.1 & 12.2 \cr 
\enddata 
\label{TabMass} 
\end{deluxetable*} 

In Table~\ref{TabMass} we list the contribution of AGNs in six different bolometric luminosity 
ranges to the total baryonic mass of galaxies and total metal mass in the simulation volume. 
This is complementary to the information presented in Figure~1 of Paper~I. 
The first column indicates the luminosity range. Columns 2, 3, and 4 indicate the number
of AGNs, the total baryonic mass of the host galaxies, and the total mass of metal, respectively,
with the corresponding percentages in columns 5, 6, and 7.
Even though low-luminosity AGNs dominate by numbers, with 76.8\% AGNs having $L<10^9L_\odot$
and 94.0\% AGNs having $L<10^{10}L_\odot$, it is the high-luminosity AGNs that contain most
of the baryonic mass and produce most of the metals. Only 1.6\% of AGNs have a luminosity in
the range $10^{11}L_\odot < L < 10^{13}L_\odot$, but they contain 77.2\% of
the baryonic mass and produce 84.4\% of the metals.



\subsection{Average Metallicity of the IGM} 
\label{sec-avgMetal} 

\begin{figure} 
\centering
\includegraphics[width = \linewidth]{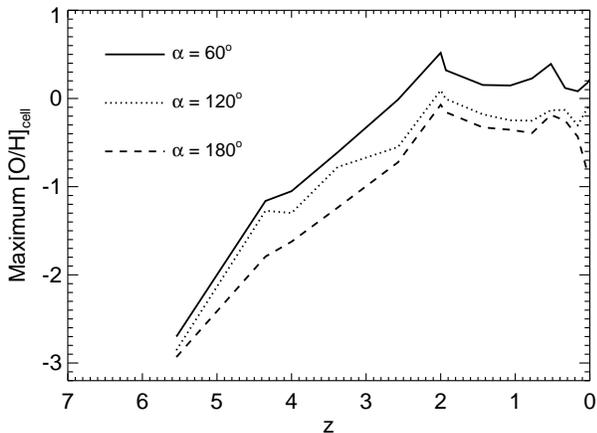}
\caption{ 
Maximum  metallicity attained by a cell 
in the whole simulation volume, 
as a function of redshift, for different outflow opening angles: 
$\alpha = 60^{\circ}$ ({\it solid}), $120^{\circ}$ ({\it dotted}), 
$180^{\circ}$ ({\it dashed}). 
}
\label{fig-lowhighMetal-alpha}
\end{figure} 

Using equations~(\ref{eq-Zcell}) and (\ref{eq-ObyHcell}), we calculated the
metal content and metallicity caused by the AGN outflows,
in a $N_{\rm ff}^3=256^3$ grid, for each run.
Figure~\ref{fig-lowhighMetal-alpha} shows the maximum metallicity 
$[{\rm O}/{\rm H}]_{\rm cell, max}$ in
the computational volume versus redshift, 
for outflows with various opening angles.
The maximum metallicity increases with time up to redshift
$z=2$. As new AGNs are formed and produce outflows, these outflows overlap
with the ones produced by earlier AGNs, causing an increase in metallicity.
After $z=2$, the value $[{\rm O}/{\rm H}]_{\rm cell, max}$ levels-off
around 0, and the variations with $z$ are no longer monotonic. At that point,
many outflows have reached the post-AGN phase. They are still expanding,
but no additional metals are being added, so the metals contained in the
outflows are diluted. This effect
competes with the addition of metals by late-forming
AGNs that are still in the active-AGN phase.

We compute the mass-weighted and volume-weighted
average metallicities induced in the IGM in our simulations. 
The averages are computed by considering only the volumes 
which are enriched by the outflows. Hence, only
cells which contain enriched particles
(among $N_{\rm ff}^3$ total in the whole simulation box, 
\S\ref{sec-computeMetal}) are included. 
In computing the metallicity $\left[ {\rm O}/{\rm H} \right]$, the $\log_{10}$ 
is taken after doing the averages over the relevant volumes. 
Mathematically, the global volume-weighted average metallicity 
in the enriched volumes is expressed as: 
\begin{eqnarray} 
\label{eq-ObyH-vol} 
& & [{\rm O/H}]_{\rm volAvg} = [\langle{\rm O/H}\rangle]_{\rm volume} \nonumber \\ 
& & \equiv  
\log_{10} \left[
\frac{\sum\limits_{\rm enriched~cells} 
\left(\frac{\nO}{\nH}\right)_{\rm cell} 
\Big/\left(\frac{\nO}{\nH}\right)_{\odot}}
{\rm Number~of~enriched~cells} \right]. 
\end{eqnarray} 
(Note: this is effectively a volume-weighted average, since all cells
have the same volume).
The global mass-weighted average metallicity in the enriched volumes 
is given by: 
\begin{eqnarray} 
& & [{\rm O/H}]_{\rm massAvg} = [\langle{\rm O/H}\rangle]_{\rm mass} \nonumber \\ 
& & \equiv
\log_{10} \left[
\frac{\sum\limits_{\rm enriched~cells}\rho_{\rm cell} 
\left(\frac{\nO}{\nH}\right)_{\rm cell} 
\Big/\left(\frac{\nO}{\nH}\right)_{\odot}} 
{\sum\limits_{\rm enriched~cells} \rho_{\rm cell}} 
\right]. 
\end{eqnarray} 
For the solar oxygen abundance we use: 
$12 + \log_{10}(\nO/\nH)_{\odot} = 8.93$ \citep{anders89}. 

\begin{figure} 
\centering
\includegraphics[width = \linewidth]{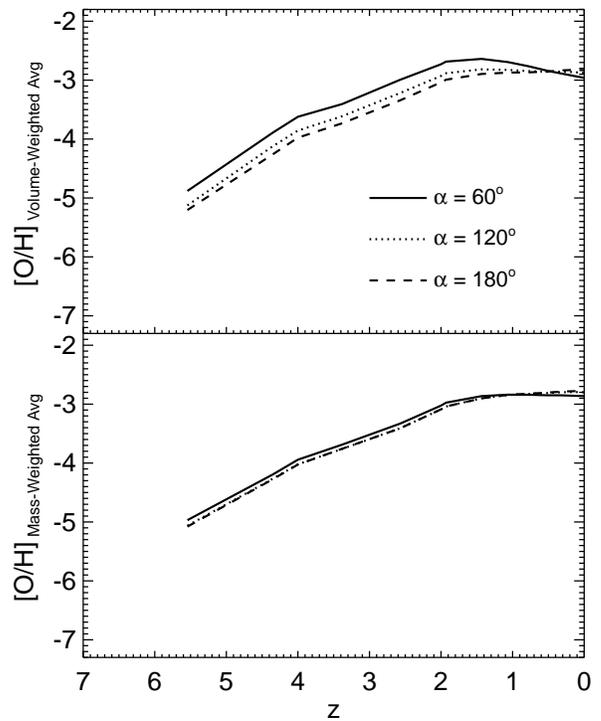}
\caption{ 
Average metallicity (volume-weighted average in the {\it top} panel, 
and mass-weighted average in the {\it bottom} panel) 
in the enriched volumes of the simulation box, 
as a function of redshift, for different outflow opening angles: 
$\alpha = 60^{\circ}$ ({\it solid}), $120^{\circ}$ ({\it dotted}), $180^{\circ}$ ({\it dashed}). 
}
\label{fig-avgMetal-alpha}
\end{figure} 

\begin{figure} 
\centering
\includegraphics[width = \linewidth]{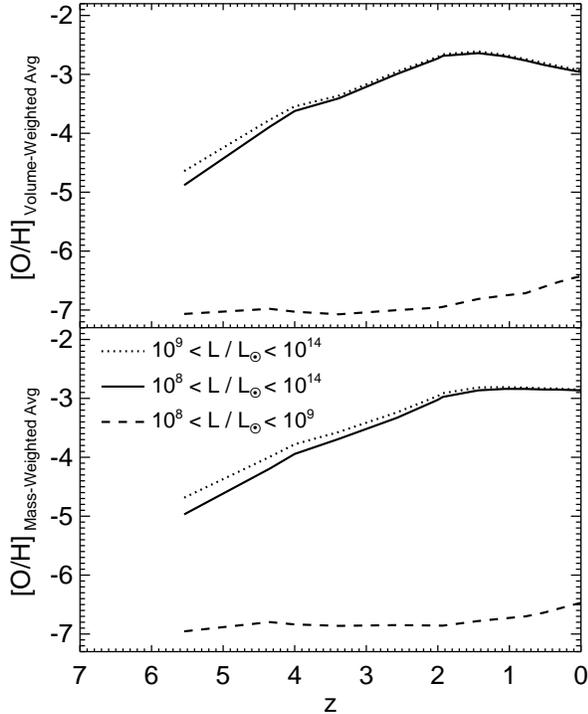}
\caption{ 
Average metallicity (volume-weighted average in the {\it top} panel, 
and mass-weighted average in the {\it bottom} panel) 
in the enriched volumes of the simulation box, 
as a function of redshift, for different bolometric luminosity limits: 
the whole population, $10^8 < L / L_{\odot} < 10^{14}$ ({\it solid}), 
only the high-luminosity sources, $10^9 < L / L_{\odot} < 10^{14}$ ({\it dotted}), 
only the low-luminosity sources, $10^8 < L / L_{\odot} < 10^9$ ({\it dashed}). 
}
\label{fig-avgMetal-Lcutoff}
\end{figure} 

The redshift evolution of $\left[ {\rm O}/{\rm H} \right]_{\rm volAvg}$ 
and $\left[ {\rm O}/{\rm H} \right]_{\rm massAvg}$ 
in the enriched volumes of the simulation box
are presented in 
Figure~\ref{fig-avgMetal-alpha} for AGN outflows with various opening angles, 
and Figure~\ref{fig-avgMetal-Lcutoff} 
for AGN populations with different bolometric luminosity limits. 
We note that the values of volume-weighted and mass-weighted averages 
in a simulation are quite similar. 
In runs A, B, C, and C1, $\left[ {\rm O}/{\rm H} \right]_{\rm volAvg}$ 
is slightly larger than $\left[ {\rm O}/{\rm H} \right]_{\rm massAvg}$ at high-$z$, 
while after a certain redshift ($z \sim 1.1$ in run A, $0.7$ in B, $0.3$ in C and C1) 
$\left[ {\rm O}/{\rm H} \right]_{\rm massAvg}$ becomes higher. 
A reverse trend is seen in run C2, where 
$\left[ {\rm O}/{\rm H} \right]_{\rm massAvg}$ is greater at high-$z$, and 
$\left[ {\rm O}/{\rm H} \right]_{\rm volAvg}$ takes over after $z \sim 0.5$. 

The whole population of AGNs 
(having bolometric luminosities between $10^8 < L / L_{\odot} < 10^{14}$) 
with most-anisotropic outflows 
(opening angle $\alpha = 60^{\circ}$, run C) induces a 
volume-averaged IGM metallicity of 
$\left[ {\rm O}/{\rm H} \right] \approx -4.9$ at $z = 5.5$, 
peaking to $\left[ {\rm O}/{\rm H} \right] \approx -2.6$ between $z \sim 1 - 1.5$, 
and then leveling off (slightly decreasing) to 
$\left[ {\rm O}/{\rm H} \right] \approx -2.9$ at the present epoch. 
Run B ($\alpha = 120^{\circ}$) and 
the isotropic outflow case ($\alpha = 180^{\circ}$, run A) 
produce quite similar average results: 
starting with $\left[ {\rm O}/{\rm H} \right] \approx -5.1$ at $z = 5.5$, 
then rising gradually, and lying more or less flat at 
$\left[ {\rm O}/{\rm H} \right] \approx -2.8$ between $z = 2 - 0$. 
There is clearly a change of slope around $z=2$, where the 
mass-averaged metallicity and volume-average metallicity level off.
Again, this can be attributed to the fact that many outflows have reached
the post-AGN phase. The resulting spreading of the metals compensates
for the injection of new metals by late-forming AGNs.

The variations of these results between runs A, B, and C are small. 
Outflows with different opening angles 
produce comparable average metallicities in the IGM. 
However, we find that more anisotropic outflows ($\alpha = 60^{\circ}$) 
causes higher average metallicity at high-$z$, 
and the trend reverses at low-$z$, 
with more isotropic outflows ($\alpha = 180^{\circ}$) 
giving larger metallicity values. At $z=0$, the average metallicities are
essentially the same for all opening angles. For the volume-average
metallicity, this was expected, since the amount of metals deposited
by AGNs is the same for all opening angles, and the enriched volume
fraction is also the same (Paper~I). For the mass-averaged metallicity,
the extent of overlaps between outflows could have made a
difference, but this is clearly not the case.

The high-luminosity AGNs ($10^9 < L / L_{\odot} < 10^{14}$, run C1) 
produce slightly higher metallicities at high-$z$; 
starting with $\left[ {\rm O}/{\rm H} \right] \approx -4.6$ at $z = 5.5$, 
and converge to similar values as the whole population (run C) by $z \sim 3$. 
The contribution of the low-luminosity AGNs ($10^8 < L / L_{\odot} < 10^9$) 
to the resultant IGM metallicity is much lower, 
$2-3$ orders orders of magnitude smaller than the luminous sources. 
Run C2 produces an average metallicity of 
$\left[ {\rm O}/{\rm H} \right] \approx -7$ at $z = 5.5$, 
which only increases to $\left[ {\rm O}/{\rm H} \right] \approx -6.4$ 
at the present epoch. We therefore
conclude that the metallicity induced in the IGM by anisotropic AGN outflows 
is greatly dominated by sources having bolometric 
luminosity $L > 10^9 L_{\odot}$, 
sources with $10^8 < L / L_{\odot} < 10^9$ have a negligible contribution. 



We compare our results with other observational and theoretical studies. 
As detailed in \S\ref{sec-intro}, 
\CIV\ and \OVI\ 
observations show that the average metallicity of the IGM 
is $Z \gtrsim 10^{-3} - 10^{-2} Z_{\odot}$ at $z \sim 2 - 3$ 
\citep{cowie95, songaila96, songaila97, carswell02, simcoe04}, 
with few indications that an IGM metallicity of $\sim 10^{-4} Z_{\odot}$ 
is already in place at $z = 5$ \citep{songaila01, ryan-weber09}. 
At low redshifts, 
a mean cosmic metallicity of $Z \gtrsim 10^{-2} - 10^{-1} Z_{\odot}$ 
is detected \citep{burles96, tripp02, danforth05}. 

Several studies and cosmological numerical simulations
have shown that ejection of metals via 
winds from galaxies can reproduce the observed IGM metal properties
\citep{tegmark93, aguirre01, scannapieco06}. 
These simulations have shown that
outflows driven by SNe or starbursts can 
produced IGM mean metallicity values consistent with observation.
In these simulations, most of the outflows originate from
dwarf and low-mass galaxies, and start at an early epoch ($6 \leq z \leq 15$). 
\citet{scannapieco02} 
found that most of the enrichment occurred 
relatively early, 
with mass-averaged cosmological metallicity 
values $10^{-3} - 10^{-1.5} Z_{\odot}$. 
\citet{oppenheimer06} obtained a global average 
metal mass fraction of $\log (Z / Z_{\odot}) \sim -2$ at $z \sim 2$. 
We found similar values for the volume-weighted and mass-weighted average, 
while \citet{oppenheimer06} obtained 
a mass-weighted average larger than the volume-weighted average. 
This is because \citet{oppenheimer06} averaged over 
the entire computational box, 
whereas we average only over the IGM volumes which are enriched in metals. 




The values of IGM metallicities we computed 
are somewhat lower compared to these observations and simulations. 
Our simulations produced an average IGM metallicity of 
$\left[ {\rm O}/{\rm H} \right] \sim -5$ at $z = 5.5$, 
then rising gradually, and lying relatively flat at 
$\left[ {\rm O}/{\rm H} \right] \approx -2.8$ between $z = 2 - 0$. 
Comparing these results with observations, we estimate that
AGN outflows contribute more than
10\% of the IGM metallicity at $z = 5$, between
16\% and 100\% at $z = 2$, and between 2\% and 16\% at $z = 0$. 
The biggest factor coming to play here is the amount of metals in the 
galaxies 
which is available to be ejected to the IGM by the outflows. 
The total galaxy baryonic mass of AGN hosts 
(galaxy population obtained from the QLF) 
is $0.0083$ of the total baryons in the simulation box at $z = 0$. 
AGNs are only hosted in a fraction of galaxies of the Universe, 
and our model tracks only the metals generated in the AGN host galaxies. 
Since we do not take into account metals generated in non-AGN host galaxies, 
over the cosmic epochs, the resulting IGM metallicities we obtained 
are smaller than ($\sim 10 \%$ of) the observed values. 


Some other simulation-based studies 
indicate that less-massive galaxies dominate in enriching the IGM at high-$z$, 
and ejection by AGN outflows accounts for only a fraction of the IGM metallicity. 
\citet{cen01} stressed that 
dwarf and subdwarf galaxies with masses $10^{6.5} - 10^{9} M_{\odot}$ 
are largely responsible for the metal pollution of the IGM 
($\sim 10^{-2} Z_{\odot}$, as observed in Ly-$\alpha$ clouds) at $z \sim 3$. 
\citet{thacker02} showed that outflows from 
galaxies with total halo masses $\lesssim 10^{10} M_{\odot}$ 
enrich $\sim 20 \%$ of the simulation volume 
with a mean metallicity of $0.3 \%$ solar at $z = 4$. 

Few direct observations of powerful outflows from AGN 
have been used to constrain the cosmological contribution of 
AGN outflows to enriching the IGM. 
Extending the observed properties of the radio galaxy MRC 1138-262 
using cosmological constraints, 
\citet{nesvadba06} estimated that powerful AGN feedback ejects 
$\sim (1 - 30) \times 10^4 Z / Z_{\odot} M_{\odot}$ Mpc$^{-3}$ of metals. 
So AGN feedback from massive galaxies may account for 
only few to $\sim 20 \%$ of all metals in the IGM at $z \sim 2$. 
Comparing the relative impact of AGN- vs.~starburst-driven winds, 
\citet{nesvadba07} did some preliminary calculations 
using the comoving number density of high-$z$ quasars, 
corrected for their duty cycle and energy injected. 
They found that the metal mass expelled by AGN winds into the IGM is smaller 
than that by starburst winds, 
concluding that AGN winds are likely not the dominant mode of IGM metal 
enrichment through cosmic time, however 
they could be a significant contributor.
Our results are consistent with such works that propagation of metals 
by AGN outflows do not account for $100 \%$ of the observed metals in the IGM. 
We conclude that ejection of metals from (those generated in) 
AGN host galaxies by AGN outflows enriches the IGM to $> 10 - 20 \%$ 
of the observed values, the number dependent on redshift. 

This could give the impression that metal enrichment by AGN-driven
outflows is a minor effect compared to metal enrichment by SNe-driven
outflows. However, one must take into account the enriched volume
fraction, which is of order $10-20\%$ 
for SNe-driven outflows \citep{sb01,madau01,bsw05,pmg07}, and over
80\%  for AGN-driven outflows (\citealt{lg05}; Paper~I). 
The final radii of AGN outflows shown here in 
Figures~\ref{rz} and \ref{rm} are significantly larger than 
the ones for SNe-driven outflows (\citealt{pmg07}, Figure~8). 
The SNe-driven outflows do not travel far from the cosmological structures 
from where they originate, 
and they are not powerful enough to reach the deepest voids. 
Thus, AGNs are likely one of the most important contributors 
to metal enrichment 
of the IGM, 
especially the low-density regions because of their ability to trigger powerful outflows.

\subsection{Enriched Volume above Metallicity Limit} 
\label{sec-enrichVol} 

\begin{figure} 
\centering
\includegraphics[width = \linewidth]{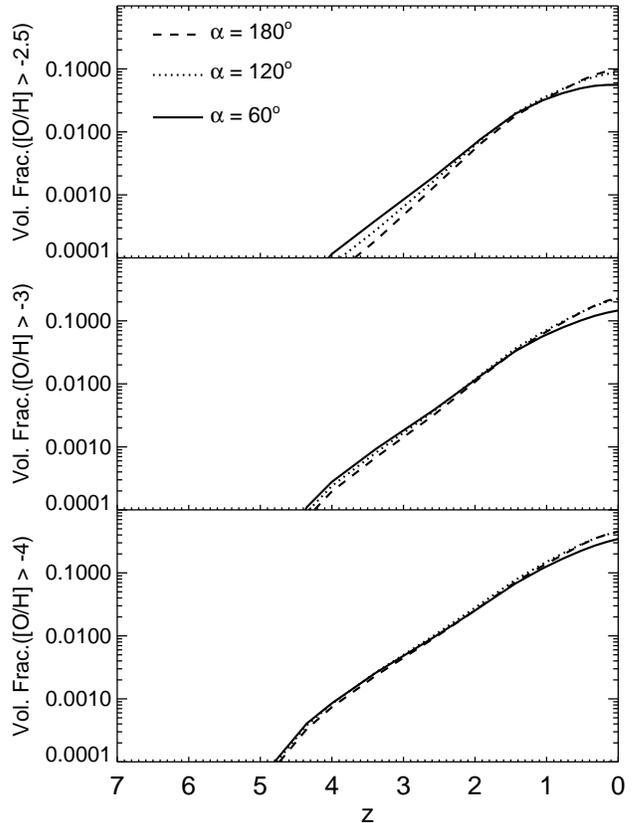}
\caption{ 
Volume fractions enriched above metallicity limits: 
$\left[ {\rm O}/{\rm H} \right] > -2.5$ in the {\it top} panel, 
$\left[ {\rm O}/{\rm H} \right] > -3$ in the {\it middle} panel, 
$\left[ {\rm O}/{\rm H} \right] > -4$ in the {\it bottom} panel, 
as a function of redshift, for different outflow opening angles: 
$\alpha = 60^{\circ}$ ({\it solid}), $120^{\circ}$ ({\it dotted}), $180^{\circ}$ ({\it dashed}). 
} 
\label{fig-enrichVol-alpha} 
\end{figure} 

\begin{figure} 
\centering 
\includegraphics[width = \linewidth]{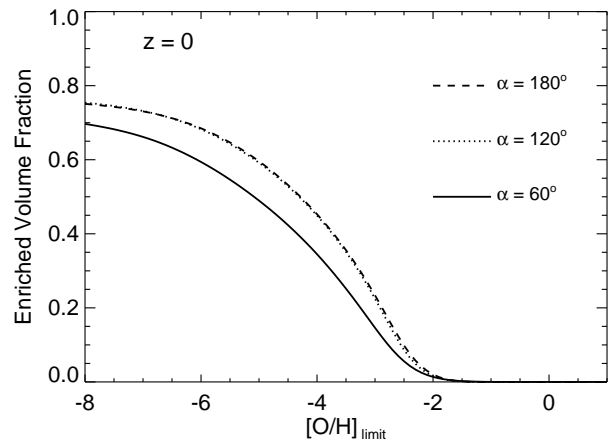} 
\caption{ 
Volume fractions enriched above a certain metallicity limit, 
$V(\left[{\rm O}/{\rm H}\right]>\left[ {\rm O}/{\rm H} \right]_{\rm limit})$, 
at the present epoch, for different outflow opening angles: 
$\alpha = 60^{\circ}$ ({\it solid}), $120^{\circ}$ 
({\it dotted}), $180^{\circ}$ ({\it dashed}). 
} 
\label{fig-enrichVol-vsMetal} 
\end{figure} 



\begin{deluxetable*}{lrcccc} 
\tablecaption{Average Metallicity and Enriched Volume Fraction at $z=0$} 
\tablewidth{0pt} 
\tablehead{ 
\colhead{Run} & 
\colhead{$\alpha$ ($^{\circ}$)} & 
\colhead{$\left[ {\rm O}/{\rm H} \right]_{\rm volAvg}$} & 
\colhead{$V_{\rm frac}(>-4)$} &
\colhead{$V_{\rm frac}(>-3)$} &
\colhead{$V_{\rm frac}(>-2.5)$} 
} 
\startdata 
A  & $180$ & $-2.830$ & 0.453 & $0.234$ & 0.098 \cr 
B  & $120$ & $-2.877$ & 0.449 & $0.225$ & 0.087 \cr 
C  & $60$  & $-2.960$ & 0.345 & $0.146$ & 0.056 \cr 
C1 & $60$  & $-2.934$ & 0.344 & $0.146$ & 0.056 \cr 
C2 & $60$  & $-6.424$ & 0.000 & $0.000$ & 0.000 \cr 
\enddata 
\label{TabVolumes} 
\end{deluxetable*} 

We calculated the volume fractions of the simulation box 
which are enriched above metallicity limits of
$\left[ {\rm O}/{\rm H} \right] = -4$, $-3$, and $-2.5$. 
In practice, we count
the number of cells in the computational volume which have
$[{\rm O/H}]_{\rm cell} > -4$, $[{\rm O/H}]_{\rm cell} > -3$, and
$[{\rm O/H}]_{\rm cell} > -2.5$
and divide by the total number ($N_{\rm ff}^3$) of cells. 
Figure~\ref{fig-enrichVol-alpha} shows the 
redshift evolution of the fractional volumes above these metallicity limits 
for AGN outflows with various opening angles. 
At high-$z$, more anisotropic outflows (lower values of $\alpha$) 
enrich slightly larger volume fractions. 
The trend reverses at low-$z$, where run B ($\alpha = 120^{\circ}$) and 
the isotropic outflow case ($\alpha = 180^{\circ}$, run A) 
enrich similar volume fractions, 
somewhat higher than the most-anisotropic 
outflows ($\alpha = 60^{\circ}$, run C). 
These enriched volume fractions are small at $z > 3$, 
then rises rapidly. 
We found that the volumes enriched above these metallicity limits by 
the high-luminosity sources ($10^9 < L / L_{\odot} < 10^{14}$, run C1) 
are $< 1 \%$ lower than that by 
the whole population of AGNs ($10^8 < L / L_{\odot} < 10^{14}$, run C). 
This is not distinguishable in this plotting scale, so we do not show them separately. 

In Table~\ref{TabVolumes} we 
list some of the main results presented in this section and
the previous one. 
The first 2 columns are the same as in Table~\ref{TabRuns}. 
The third column gives the resulting volume-weighted average metallicity 
($\left[ {\rm O}/{\rm H} \right] _{\rm volAvg}$) at the present epoch.
The fourth, fifth, and sixth columns give the volume fraction enriched
above metallicity of $\rm[O/H]=-4$, $-3$, and $-2.5$ at
the present epoch, respectively. While the average metallicity is quite
insensitive to the opening angle (varying from $-2.83$ to $-2.96$),
There is a significant dependence of the volume fraction on the opening angle,
with larger opening angles resulting in larger values. At larger opening 
angles, there is more overlap between outflows, resulting in larger
values of the metallicity in the regions where overlap occurs.
The low-luminosity AGNs ($10^8 < L / L_{\odot} < 10^9$, run C2) 
do not enrich any volume to any of these metallicity limits 
($\left[ {\rm O}/{\rm H} \right] > -4$) at any time. 
This is not surprising, since the average metallicity it produces 
is only $\left[ {\rm O}/{\rm H} \right] \approx -6.4$ at present. 

We compare these results with some related contemporary studies. 
\citet{scannapieco02} performed Monte Carlo cosmological simulations to track 
the time evolution of SN-driven metal-enriched outflows from early galaxies, 
finding that up to $30 \%$ volume of the IGM 
is enriched to above $10^{-3} Z_{\odot}$ at $z = 3$, 
making the enrichment biased to the areas near the 
starbursting galaxies themselves. 
\citet{oppenheimer06} found that their resulting volume fraction having 
$\left[ {\rm Z}/{\rm H} \right] > -3$ reaches 
$\sim 10 \%$ at $z = 2$. 
\citet{oppenheimer09} calculated that the volume filling factor of metals 
increases between $z = 8$ and $z=5$, 
reaching $\sim 1 \%$ for $Z > 10^{-3} Z_{\odot}$ by $z = 5$. 
The volumes we obtained enriched above given metallicity limits 
are smaller than these studies, 
for reasons same as that discussed in \S\ref{sec-avgMetal}. 
We only eject the metals produced in AGN host galaxies. 
At high redshifts, AGN-driven 
outflows enrich smaller volume fraction of the IGM 
above a certain metallicity than SNe-driven
outflows which transport metals from less-massive galaxies as well. 

The continuous behavior of the enriched IGM volumes as a 
function of metallicity 
is presented in Figure~\ref{fig-enrichVol-vsMetal}. 
It shows the fractional volume enriched above a certain metallicity limit 
at $z = 0$, caused by AGN-driven outflows with various opening angles. 
We see that
the more isotropic outflows ($\alpha = 120^{\circ}$ and $180^{\circ}$) 
enrich almost the same volume fractions above a given metallicity, 
$5 - 10 \%$ higher than the most-anisotropic outflows ($\alpha = 60^{\circ}$). 
More than $60 \%$ of the volume is enriched
above $\left[ {\rm O}/{\rm H} \right] > -6$, but only
and $1 - 2 \%$ of the volume is 
enriched above $\left[ {\rm O}/{\rm H} \right] > -2$. 
The curves we obtained in Figure~\ref{fig-enrichVol-vsMetal} 
are qualitatively similar to the one
shown in Figure~5 of \citet{thacker02}. 
Even though the scales and redshift are different (and they
consider SNe-driven outflows), the trends are similar. At high abundances, 
there is a rapid increase in enriched volume fraction
with decreasing metallicities, while at lower abundances
the enriched volume fraction tends to get flatter.

\subsection{Density - Enrichment Correlation} 
\label{sec-DensityEnrich} 

We compute the mean metallicity of the IGM at a given
density, $[{\rm O}/{\rm H}]_{\rho}$, 
within the range $-3 \leq \log (\rho/\bar{\rho}) \leq 4$, where
$\bar\rho=3\Omega_0H_0^2(1+z)^3/8\pi G$ is the mean
density of our simulation box at redshift $z$.
We divide the density range in bins of size 0.1~dex, and identify
the cells 
which are enriched and 
whose total matter density $\rho_{\rm cell}$
occur within each bin (see \S\ref{sec-computeMetal}). 
We average the gas density $\rho_g=\rho_{\rm cell}\Omega_{b,0}/\Omega_0$
and metal density $\rho_{\rm metal}$ within each bin to get the
bin-averaged values $\rho_{\rm g, bin}$, $\rho_{\rm metal, bin}$,
and $Z_{\rm cell,bin}=\rho_{\rm metal, bin}/\rho_{\rm g, bin}$.
We then calculate the mean metallicity
$\rm [O/H]_\rho=\log(\nO/\nH)_\rho-\log(\nO/\nH)_\odot$ inside each bin,
where $\log(\nO/\nH)_\rho$ is calculated using equation~(\ref{eq-ObyHcell}).

\begin{figure*} 
\centering 
\includegraphics[width = 0.95 \linewidth]{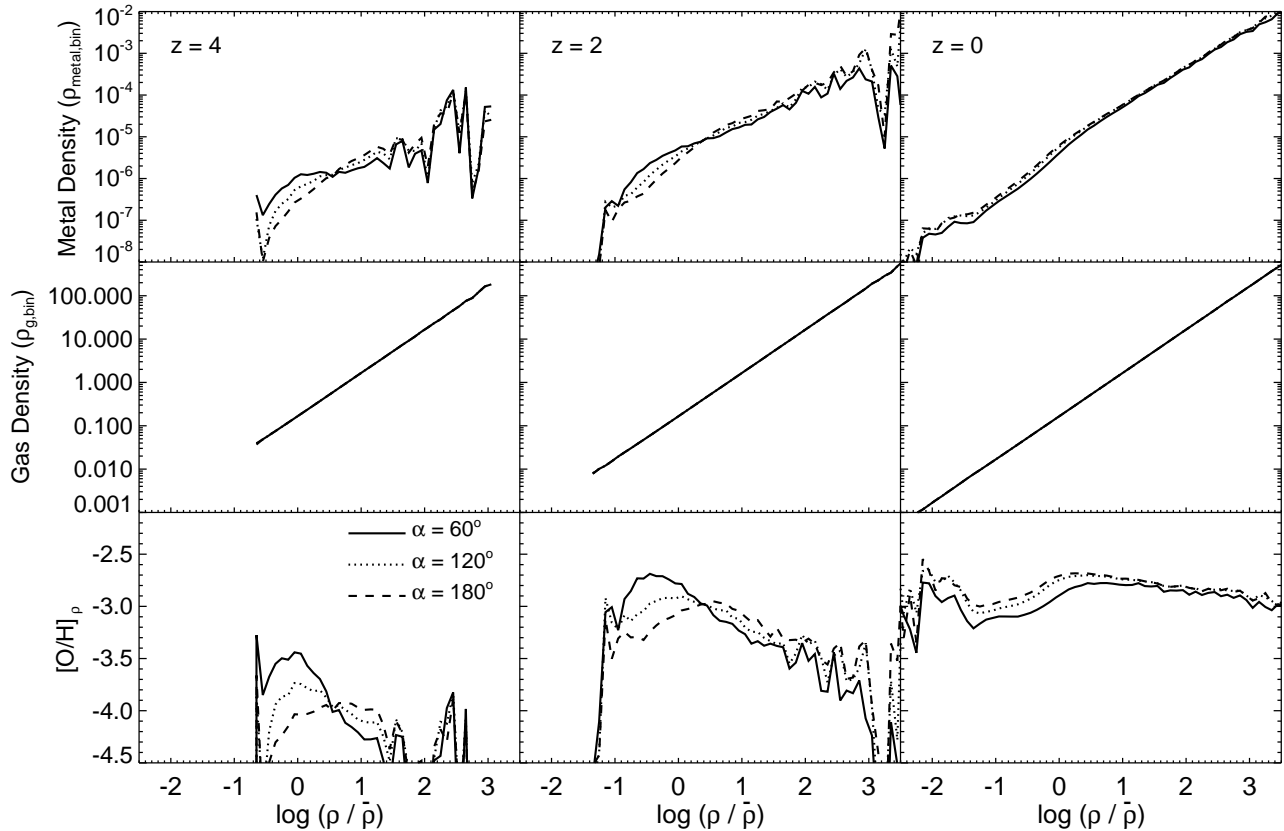} 
\caption{ 
Metallicity induced by AGN outflows  
as a function of the total density of the IGM, 
at redshifts $z = 4$ (left column), $z = 2$ (middle column) 
and $z = 0$ (right column). 
{\it top}: metal density, {\it middle}: gas density, 
{\it bottom}: metallicity 
(logarithm of ratio of the top to the middle panel, minus logarithm of the
ratio for the solar values).
The line types indicate different outflow opening angles: 
$\alpha = 60^{\circ}$ ({\it solid}), $120^{\circ}$ ({\it dotted}), 
$180^{\circ}$ ({\it dashed}). 
} 
\label{fig-density-Metals} 
\end{figure*} 

The evolution of the resulting metallicity as a function of IGM density 
(total density, including both baryons and dark matter) 
is shown in Figure~\ref{fig-density-Metals}, 
at redshifts $z = 4$, 2, and 0, in the columns from left to right.
As the simulation proceeds, overdense regions collapse while voids
become deeper, increasing the range of densities present in the
simulation volume. Also, with time the population of
outflows cover a larger fraction of the simulation volume, causing
densities of a larger range (denser filaments and less dense void
regions) to become enriched. As a result, the range of densities
($x$-axis) covered by each curve increases from $z=4$ to $z=0$. 
At a given total density, 
the metal density ($\rho_{\rm metal, bin}$, top panels) increases 
with decreasing redshift. 
This happens because with time new outflows are born, 
which continuously add more metals to the IGM volumes. 
The middle panels show the gas density ($\rho_{\rm g, bin}$) of the
enriched IGM volumes, which by construction follow the total density at 
all times (we assume $\rho_g=\Omega_{b,0}\rho/\Omega_0$). 

The bottom panels of Figure~\ref{fig-density-Metals} show 
the metallicity $[{\rm O}/{\rm H}]_{\rho}$, which grows with time 
steadily (at a given IGM density). 
This is consistent with the trend found by \citet{oppenheimer06} 
in their Figure~10. 
We find some interesting trends in the metallicity-density correlations 
as well as some differences between outflows with different opening angles. 
At higher redshifts ($z = 4$ and $2$) 
the underdense regions [$\log(\rho/\bar{\rho})<0$]
are enriched to higher metallicities, 
and the induced metallicity decreases with increasing IGM density. 
This trend is more prominent with increasing anisotropy of the outflows. 
It is very clear in the bottom-left and bottom-middle panels 
in the plot for run C ($\alpha = 60^{\circ}$). 
At $z=2$, the metallicity still decreases with increasing density
in overdense regions [$\log(\rho/\bar\rho)>0$] for all opening angles.
At $z = 0$, the metallicity is essentially flat over all IGM densities, 
for outflows with any opening angle. At that redshift, the enriched volume
fraction exceeds 80\% (Paper~I), hence all regions, at all densities,
are enriched in metals, except the deepest voids [$\log(\rho/\bar\rho)<-2.4$]
located far from any AGN. Initially, more anisotropic outflows preferentially
enrich low-density regions, but this trend gets eventually washed-out as
the enriched volume fraction approaches unity. 

Our trend of larger metallicity induced in the IGM at low densities 
for the more anisotropic outflows at high-$z$ 
is opposite from the results of \citet{oppenheimer06}, 
who found a steady increase of metallicity with overdensity. 
The reason for this discrepancy is the anisotropy of the outflows in our model. 
The AGN outflows propagate along the direction of least resistance around a density peak 
(\S\ref{sec-BasicAlgo}), 
and preferentially enrich the low-density regions to higher metallicities at early epochs. 
In our simulations, the isotropic outflow case ($\alpha = 180^{\circ}$) produces 
uniform metallicity at all densities being enriched. 

As we have detailed in \S\ref{sec-intro}, 
metal enrichment of the IGM is found to be highly inhomogeneous, 
strongly dependent on local density and redshift. 
The first detection of \OVI\ in the low-density IGM at high redshift 
was reported by \citet{schaye00}, 
who performed a pixel-by-pixel search for \OVI\ absorption 
in quasar spectra over the redshift range $z = 2 - 4.5$. 
They detected \OVI\ at $2 \leq z \leq 3$ down to 
$\tau_{\rm HI}^{\phantom1} \sim 10^{-1}$, 
showing that the IGM is enriched down to much lower overdensities, 
and did not detect \OVI\ at $z > 3$. 
Enrichment by powerful AGN outflows at high-$z$ 
(following our model prescriptions) 
can naturally explain observations of such metal-enriched underdense regions. 
Providing support to such a AGN-driven IGM enrichment scenario, 
\citet{khalatyan08} found, using SPH simulations, that 
without AGN feedback metals are confined to vicinities of galaxies, 
underestimating the observed metallicity of the IGM at overdensities 
$\lesssim 10$ (the underdense IGM).

\subsection{Diffuse and Dense IGM} 
\label{sec-DiffDen} 

\begin{figure} 
\centering
\includegraphics[width = \linewidth]{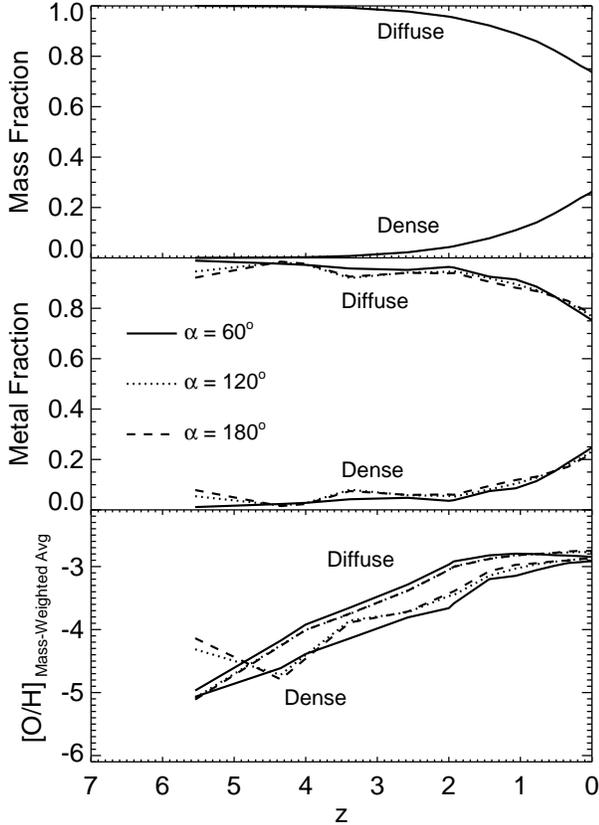}
\caption{ 
Mass fractions and metallicity for 
the diffuse ($\rho / \bar{\rho} < 100$, upper curves) 
and the dense ($\rho / \bar{\rho} \geq 100$, lower curves) 
regions of the simulation volume, as a function of redshift.
Total mass fraction in the {\it top} panel, 
metal mass fraction in the {\it middle} panel, 
and mass-weighted average metallicity within the enriched volumes 
in the {\it bottom} panel). The line types correspond
to different outflow opening angles: 
$\alpha = 60^{\circ}$ ({\it solid}), $120^{\circ}$ ({\it dotted}), 
$180^{\circ}$ ({\it dashed}), as indicated. 
} 
\label{fig-diffuseDense} 
\end{figure} 

To complete the analysis of our simulations, we divide the
whole simulation volume into two components: 
the {\it diffuse\/} IGM having density $\rho / \bar{\rho} < 100$
and the {\it dense\/} IGM with density $\rho / \bar{\rho} \geq 100$. 
We compute the mass-weighted average metallicity in the enriched volumes 
for the diffuse and dense IGM using, 
\begin{eqnarray} 
& & \left[\frac{\rm O}{\rm H} \right]_{\rm massAvg,~diffuse} = \nonumber \\ 
& & \log_{10}\left[ 
\frac{\sum\limits_{{\rm enriched~cells},\,\rho/\bar\rho<100}
\left\{\rho_{\rm cell}\left(\frac{\nO}{\nH}\right)_{\rm cell}/
\left(\frac{\nO}{\nH}\right)_{\odot} \right\}} 
{\sum\limits_{{\rm enriched~cells},\,\rho/\bar\rho<100}
\rho_{\rm cell}}\right], 
\end{eqnarray} 

\begin{eqnarray} 
& & \left[ \frac{\rm O}{\rm H} \right]_{\rm massAvg,~dense} = \nonumber \\ 
& & \log_{10}
\left[ \frac{\sum\limits_{{\rm enriched~cells},\,\rho/\bar\rho\geq100}
\left\{\rho_{\rm cell}\left(\frac{\nO}{\nH} \right)_{\rm cell}/
\left(\frac{\nO}{\nH} \right)_{\odot} \right\} } 
{\sum\limits_{{\rm enriched~cells},\,\rho/\bar\rho\geq100}
\rho_{\rm cell}} \right]. 
\end{eqnarray} 

Figure~\ref{fig-diffuseDense} shows the redshift evolution of various 
quantities 
(mass fraction, metal fraction, 
and mass-weighted average metallicity within the enriched volumes) 
in the diffuse and dense IGM, for different outflow opening angles. 
In each panel, the upper set of curves is for the 
diffuse volumes ($\rho / \bar{\rho} < 100$), 
and the lower set of curves denote the dense volumes 
($\rho / \bar{\rho} \geq 100$). 

As time goes on, the mass fraction in the dense volumes increases 
as condensed structures grow within the simulation volume, 
by accreting matter from the diffuse IGM, whose mass fraction 
then decreases. 
We find a similar trend of the metal fraction, 
the diffuse regions containing more metals than the dense regions. 
Consequently, as seen in the lower panel of Figure~\ref{fig-diffuseDense}, 
the average metallicity of the diffuse volumes is larger than 
that of the dense volumes. 
These trends are contrary to the ones found by
\citet{oppenheimer06} (their Fig.~7).
There are two reasons for this discrepancy. First, as we explained before, 
they have included all galaxies in their simulations, while we only
include AGN-host galaxies. Second, they consider SNe-driven outflows
only while we consider AGN-driven outflows. SNe-driven outflows achieve
enriched volume fractions of order 20\% compared to over 80\% for
AGN-driven outflows, which are therefore much more efficient in enriching
low-density regions.

\subsection{Different Initial Location Scheme: More Luminous AGN in Denser Peaks} 
\label{sec-InitLoc} 

\begin{figure*} 
\centering
\includegraphics[width = 0.8 \linewidth]{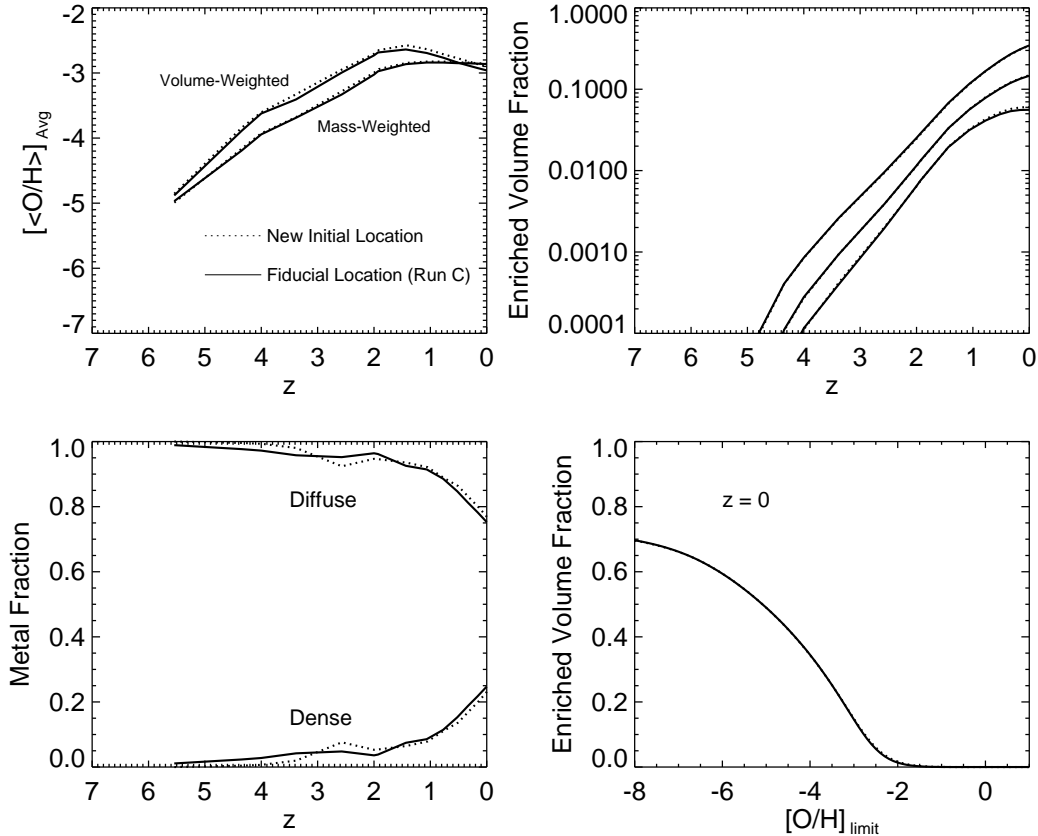}
\caption{ 
Comparison of results with the new AGN initial location prescription vs.~the fiducial scheme. 
The older results (run C) are shown as {\it solid} lines. 
The new run is shown by {\it dotted} lines, lying very close to the old run in each case. 
{\it Top-left}: Volume-weighted average (upper curves), 
and mass-weighted average (lower curves) metallicity, in the enriched volumes of the simulation box. 
{\it Top-right}: 
Curves, from top to bottom, correspond to volume fractions enriched above metallicity limits: 
$[{\rm O/H}] > -4$, $[{\rm O/H}] > -3$, and $[{\rm O/H}] > -2.5$, respectively. 
{\it Bottom-left}: Metal mass fraction in the diffuse (upper curves), 
and the dense (lower curves) regions of the simulation volume. 
{\it Bottom-right}: Volume fraction enriched above a certain metallicity limit at the present epoch. 
} 
\label{fig-InitLoc} 
\end{figure*} 

The results presented in the previous subsections 
are based on the 5 simulation runs 
done using the same fiducial AGN initial location prescription as in our Paper I, 
as mentioned in \S\ref{sec-BasicAlgo}. 
Restating the method: we spatially locate the new AGNs 
born during a timestep interval at the local density peaks, 
selected randomly among all the peaks in the cosmological volume at that timestep of the simulation. 
(Runs H and I of our Paper I considered density bias and clustering in locating the AGNs.) 

In order to test the robustness of this fiducial AGN initial location scheme 
w.r.t.~AGN-luminosity--peak-density correlations, 
we performed a test simulation run by implementing a new prescription. 
In this run, 
after selecting the peaks at each timestep to locate the new AGNs, 
the peaks were sorted according to their densities 
and the AGNs were sorted according to their luminosities. 
Then each AGN was assigned a peak such that 
the highest-luminosity AGN is located in the highest-density peak, and so on. 
This method was repeated at all the timesteps. 

The results of this test run came out to be almost similar to our 
corresponding run with the fiducial location scheme. 
Figure~\ref{fig-InitLoc} shows a comparison of the results between the two runs. 
The top-left panel shows that the mass-weighted average metallicity is undistinguishable, 
and the new run produced a slightly higher volume-weighted average metallicity. 
The volume fractions enriched above metallicity limits are undistinguishable 
(from the top-right and bottom-right panels). 
The quantity that showed the relatively largest difference 
is the metal mass fraction in the diffuse and the dense regions 
of the simulation volume, shown in the bottom-left panel. 
At the same time, the extent of this slight difference reduces at lower redshifts ($z < 2$). 

Overall there is a $\sim 1\%$ difference. 
This shows that our results are not sensitive to correlating 
the AGN-luminosity with the peak-density in our AGN initial location prescription. 
In our model, at every timestep, we use the observed QLF 
to calculate the number of AGNs born during that timestep, 
and we locate these AGNs in density peaks 
that are available at that timestep. 
Such a technical implementation averages out 
any effect arising from correlating the AGN-luminosity with the peak-density. 



\section{SUMMARY AND CONCLUSION} 
\label{sec-conclusion} 

We have computed the metallicity of the IGM induced by outflows 
of the cosmological population of AGNs over the age of the Universe. 
In Paper~I, 
we developed a semi-analytical model for anisotropic
AGN outflows expanding along the path of least resistance, 
and implemented it into $N$-body cosmological simulations. 
In this paper, we have incorporated a detailed metal enrichment technique 
in our AGN outflow model from Paper~I, 
and have simulated the metal-enrichment of the IGM. 
Our enrichment prescription consists of 
estimating the masses of metals transported with the outflows 
from the host galaxies, 
and distributing the metals to the IGM volumes being intercepted by 
the outflows. 

During their expansion, 
the AGN outflows carry with them metals generated within the host 
galaxy (by stars), 
and spread the metals to the broader IGM. 
We distinguish 2 regions of stellar populations as the sources of 
metal generation. 
For the stars near the AGN, we consider a central metal abundance 
of $5 Z_{\odot}$ 
(consistent with observed super-Solar metallicity of broad-line regions). 
We further take into account that AGN outflows can entrain some enriched ISM 
of the greater galaxy and spread it to the IGM. 
For stars located away from the central AGN within the rest of the galaxy, 
we use the redshift-dependent mass-metallicity relation of galaxies 
derived from observations. 
Enrichment of the IGM is simulated by 
distributing the metals transported by the outflows to the 
intergalactic gas 
intercepted by them. 
Each outflow is assumed to impart all of the stripped metals from its host galaxy to the IGM 
over the active-AGN lifetime, depositing metals at an uniform rate in time. 

Using this algorithm,
we simulated the propagation of AGN-driven outflows carrying metals, 
along with the growth of large-scale structures, 
in a cosmological volume of size $(128~h^{-1}$ Mpc $)^3$ 
in a $\Lambda$CDM universe, 
and analyzed the resulting metal-enrichment of the IGM. 
We performed a series of 5 simulations by varying the outflow opening angle 
and the AGN population limiting luminosities. 
The main results and conclusions from our study are in the following. 

1) In our simulation box of comoving volume ($128~h^{-1}$ Mpc)$^3$, 
the total baryonic mass of all the AGN host galaxies 
is $3.18 \times 10^{14} M_{\odot}$, 
and the total mass of metals available to be distributed to the IGM 
(by the outflows) is $1.18 \times 10^{12} M_{\odot}$, 
at the present epoch. 
The ratio of total baryonic mass of galaxies 
over that of the simulation box is $0.00828$. 
The total metal mass is a fraction $0.00370$ of the baryonic mass in galaxies, 
and $3.07 \times 10^{-5}$ of the baryonic mass simulation box. 
The masses of the whole AGN population ($10^8 < L / L_{\odot} < 10^{14}$) 
is completely dominated by the high-luminosity 
sources ($10^9 < L / L_{\odot} < 10^{14}$). 
The low-luminosity AGNs ($10^8 < L / L_{\odot} < 10^9$) 
have a very small contribution to the total masses, 
at $z = 0$ they comprise of $1.7 \%$ of the total galaxy mass 
and $0.1 \%$ of the total metal mass. 
Even though high-luminosity AGNs are much less numerous
than low-luminosity ones, 
they really dominate the mass budget, 
and form the dominant source of metals to be distributed to the IGM. 

2) The values of volume-weighted and mass-weighted 
average metallicity induced in the IGM are quite similar, in each simulation. 
The whole population of AGNs 
(having bolometric luminosities between $10^8 < L / L_{\odot} < 10^{14}$) 
induces an average IGM metallicity of 
$\left[ {\rm O}/{\rm H} \right] \approx -5$ at $z = 5.5$, which
then rises gradually, and flattens at a value
$\left[ {\rm O}/{\rm H} \right] \approx -2.8$ between $z = 2 - 0$. 
Outflows with different opening angles 
produce comparable average metallicities in the IGM. 
However, more anisotropic outflows ($\alpha = 60^{\circ}$) 
causes slightly higher average metallicity at high-$z$, 
and the trend reverses at low-$z$, 
with more isotropic outflows ($\alpha = 180^{\circ}$) 
giving larger metallicity values. 
The high-luminosity AGNs ($10^9 < L / L_{\odot} < 10^{14}$) 
produce slightly higher metallicities at high redshift, 
starting with $\left[ {\rm O}/{\rm H} \right] \approx -4.6$ at $z = 5.5$, 
and converging to similar values as the whole population by $z \sim 3$. 
The contribution of the low-luminosity AGNs ($10^8 < L / L_{\odot} < 10^9$) 
to the resultant IGM metallicity is much lower, 
$2-3$ orders of magnitude smaller than the luminous sources. 
We conclude that the metallicity induced in the IGM 
by anisotropic AGN outflows 
is greatly dominated by sources having bolometric 
luminosity $L > 10^9 L_{\odot}$, 
sources with $10^8 < L / L_{\odot} < 10^9$ having a negligible contribution. 

3) Observational studies indicate that the average metallicity of the IGM 
is $Z \gtrsim 10^{-3} - 10^{-2} Z_{\odot}$ at $z \sim 2 - 3$ 
\citep[e.g.,][]{songaila96, simcoe04}, 
with few indications that an IGM metallicity of $\sim 10^{-4} Z_{\odot}$ 
is already in place at $z = 5$ \citep{songaila01, ryan-weber09}. 
At low redshifts, 
a mean cosmic metallicity of $Z \gtrsim 10^{-2} - 10^{-1} Z_{\odot}$ 
is detected \citep{tripp02, danforth05}. 
Some cosmological numerical simulations 
\citep[e.g.,][]{scannapieco02, oppenheimer06} 
have deduced IGM mean metallicity values consistent with observations, 
taking into account metals ejected by outflows driven by SNe or starbursts, 
most of which come from low-mass galaxies, starting from early epochs ($6 \leq z \leq 15$). 
Our computed results of average IGM metallicity 
($\left[ {\rm O}/{\rm H} \right] = -5$ at $z = 5.5$, 
then rising to $\left[ {\rm O}/{\rm H} \right] = -2.8$ at $z = 2 - 0$) 
are somewhat smaller compared to these observations and simulations. 
The reason is that our model tracks only the metals produced and ejected from 
galaxies hosting AGNs, which is a fraction of the cosmological 
population of galaxies in the Universe, 
(we do not take into account metals generated in non-AGN galaxies), 
and that cannot account for $100 \%$ of the metals observed in the IGM. 
We conclude that ejection of metals from 
AGN host galaxies by AGN outflows enriches the IGM to $> 10 - 20 \%$ 
of the observed values, the number dependent on redshift. 
Some other simulation-based work 
indicates that less-massive galaxies dominate the enrichment 
the IGM at high-$z$ 
\citep{cen01, thacker02}, 
supporting the scenario that 
ejection by AGN outflows accounts for only a fraction of the IGM metallicity. 
From direct observations of powerful outflows from AGN, 
\citet{nesvadba06} estimated that powerful AGN feedback from massive galaxies 
may account for only few to $\sim 20 \%$ of all metals in the
 IGM at $z \sim 2$. 
Comparing the relative impact of AGN- vs.~starburst-driven winds, 
\citet{nesvadba07} concluded that AGN winds are likely not the 
dominant mode of IGM metal 
enrichment through cosmic time, however they could be a significant 
contributor. 

4) At high-$z$, more anisotropic outflows (lower values of $\alpha$) 
enrich slightly larger volume fractions. 
The trend reverses at low-$z$, where the runs with 
$\alpha = 120^{\circ}$ and $180^{\circ}$ enrich similar volume fractions, 
somewhat higher than the most-anisotropic outflows ($\alpha = 60^{\circ}$). 
These enriched IGM volume fractions above certain metallicity limits are small at $z > 3$, 
then rises rapidly to the following values at the present epoch: 
$5.8 - 10 \%$ of the volume enriched to $\left[ {\rm O}/{\rm H} \right] > -2.5$, 
$14.5 - 24 \%$ volume to $\left[ {\rm O}/{\rm H} \right] > -3$, 
and 
$33.8 - 45 \%$ volume to $\left[ {\rm O}/{\rm H} \right] > -4$. 
The more isotropic outflows ($\alpha = 120^{\circ}$ and $180^{\circ}$) 
are seen to enrich almost the same volume fractions above a given metallicity, 
$5 - 10 \%$ higher than the most-anisotropic outflows ($\alpha = 60^{\circ}$). 
At $z = 0$, $> 60 \%$ of the volume is enriched to $\left[ {\rm O}/{\rm H} \right] > -6$, 
and $1 - 2 \%$ volume has $\left[ {\rm O}/{\rm H} \right] > -2$. 
Our results of fractional volumes enriched as a function of metallicity limits 
are consistent with that of \citet{thacker02}. 


5) Examining the correlations of the induced enrichment with the IGM density, 
we find that 
the metallicity $[{\rm O}/{\rm H}]_{\rho}$ grows with time steadily (at a given IGM density). 
This is consistent with the trend found by \citet{oppenheimer06}. 
At higher redshifts ($z \geq 2$), for more anisotropic outflows, 
there is a prominent gradient of the induced enrichment, 
the metallicity decreasing with increasing IGM density, 
making the underdense ($\rho < \bar{\rho}$) regions enriched to higher metallicities. 
This trend is more prominent with increasing anisotropy of the outflows. 
The metallicity gradient (vs.~density) decreases as the outflows 
become more isotropic (larger values of $\alpha$). 
At $z = 0$, the metallicity is more or less flat over all IGM densities, 
for outflows with any opening angle. 

In Paper~I, we had found that 
increasingly anisotropic AGN outflows preferentially enrich lower-density volumes of the IGM. 
In the present work, we complement these results by showing that 
more anisotropic  outflows enrich the underdense IGM to higher metallicities at high-$z$. 
Evidence of substantial metallicity in underdense regions of the IGM 
at $z \sim 4$ \citep[e.g.,][]{schaye03} 
requires a strong mechanism of spreading metals widely 
into low-density regions, at early cosmic epochs.
SNe-driven outflows reach enriched volume fractions of order 20\% only,
and therefore cannot enrich low-density regions easily.
Enrichment by powerful AGN outflows at high-$z$ 
(following our model prescriptions) 
can naturally explain observations of such metal-enriched underdense regions.

6) We found that that our results are not sensitive to correlating 
the AGN-luminosity with the peak-density in our AGN initial location prescription. 
In other words, in our model 
when locating the AGNs born at a timestep 
in the density peaks available at that timestep, 
it does not make a difference 
if the more luminous AGNs are located in the denser peaks, 
or whether the AGNs are located in the available peaks randomly.

7) We found that the anisotropy of the AGN outflows affect 
most of the final results only marginally, except at higher-$z$.  
Outflows with different opening angles produced comparable 
volume-weighted and mass-weighted average metallicities in the IGM at all redshifts. 
The IGM volume fractions enriched above metallicity limits 
is larger for more anisotropic outflows at high-$z$, 
while at $z < 1.5$ more isotropic outflows enriched a larger fraction. 
At $z \geq 2$, more anisotropic outflows enriched the underdense IGM to higher metallicities.

\acknowledgments 

We thank Roberto Maiolino, Benjamin Oppenheimer,
and Matthew Pieri for useful correspondence. 
The subroutine that calculates 
the direction of least resistance was written by C\'edric Grenon. 
All calculations were performed at the Laboratoire 
d'astrophysique num\'erique, Universit\'e Laval. 
We thank the Canada Research Chair program and NSERC for support.

\appendix

\section{THE METALLICITY OF THE GALAXIES}

Equation~(\ref{eq-metal}) provides the oxygen abundance versus 
mass and redshift. To calculate a total metallicity $Z_G$, we need to make
assumptions about the abundances of the other elements. The metallicity $Z_G$
is given by
\begin{equation}
\label{A1}
Z_G={\displaystyle\sum_{Z>2}n^{\phantom1}_Z\mu^{\phantom1}_Z\over
\displaystyle \nH\mu^{\phantom1}_{\rm H}
+n^{\phantom1}_{\rm He}\mu^{\phantom1}_{\rm He}
+\sum_{Z>2}n^{\phantom1}_Z\mu^{\phantom1}_Z}\,,
\end{equation}

\noindent where $n^{\phantom1}_Z$ and 
$\mu^{\phantom1}_Z$ are the number density and atomic mass
for element $Z$, respectively, and the sum is over all metals, that is
all elements excluding hydrogen and helium. We rewrite equation~(\ref{A1})
as
\begin{equation}
\label{A2}
Z_G={\displaystyle \nO\sum_{Z>2}
{n^{\phantom1}_Z\over \nO}\,\mu^{\phantom1}_Z\over
\displaystyle \nH\mu^{\phantom1}_{\rm H}
+n^{\phantom1}_{\rm He}\mu^{\phantom1}_{\rm He}
+\nO\sum_{Z>2}
{n^{\phantom1}_Z\over \nO}\,\mu^{\phantom1}_Z}
={\nO\mu^{\phantom1}_{Z,{\rm O}}\over
\nH\mu^{\phantom1}_{\rm H}
+n^{\phantom1}_{\rm He}\mu^{\phantom1}_{\rm He}
+\nO\mu^{\phantom1}_{Z,{\rm O}}}\,,
\end{equation}

\noindent where $n^{\phantom1}_{\rm O}$ is the number density of
oxygen, and $\mu^{\phantom1}_{Z,{\rm O}}$ is the {\it equivalent oxygen
atomic mass}, defined by
\begin{equation}
\label{A3}
\mu^{\phantom1}_{Z,{\rm O}}
\equiv\sum_{Z>2}{n^{\phantom1}_Z\over \nO}\,\mu^{\phantom1}_Z\,.
\end{equation}

\noindent We assume that, when oxygen is produced, other
metals are produced at the same rate, so the ratios
$n^{\phantom1}_Z/n^{\phantom1}_{\rm O}$ remain constant. Of course, the
actual ratios will depend on the relative importance of Type~Ia and Type~II
supernovae, among other things, but this is a convenient approximation.
We calculated $\mu^{\phantom1}_{Z,{\rm O}}$ using the solar-system
abundances $n^{\phantom1}_Z$ given by \citet{cox00}.
We only consider metals having an abundance $\log\,n^{\phantom1}_Z > 5$ 
(where for oxygen, $\log\,n^{\phantom1}_{\rm O} = 8.93$), 
which includes the elements: 
C, N, O, Ne, Na, Mg, Al, Si, P, S, Cl, Ar, K, Ca, Cr, Mn, Fe, and Ni.
The resulting value is 
$\mu^{\phantom1}_{Z,{\rm O}}=31.6238$.\footnote{The main contributors are
oxygen, iron, and carbon.}

We then rewrite equation~(\ref{A2}) as
\begin{equation}
\label{A4}
Z_G={\nO\mu^{\phantom1}_{Z,{\rm O}}\over
\displaystyle
\nH\left(\mu^{\phantom1}_{\rm H}+{n^{\phantom1}_{\rm He}\over \nH}
\,\mu^{\phantom1}_{\rm He}\right)+\nO\mu^{\phantom1}_{Z,{\rm O}}}
={\nO\mu^{\phantom1}_{Z,{\rm O}}\over\nH\mu^{\phantom1}_{\rm pr,H}
+\nO\mu^{\phantom1}_{Z,{\rm O}}}\,,
\end{equation}

\noindent where $\mu^{\phantom1}_{\rm pr,H}$ is the {\it equivalent hydrogen
atomic mass}, defined by
\begin{equation}
\label{A5}
\mu^{\phantom1}_{\rm pr,H}
\equiv\mu^{\phantom1}_{\rm H}
+{n^{\phantom1}_{\rm He}\over \nH}
\,\mu^{\phantom1}_{\rm He}\,.
\end{equation}

\noindent We use the primordial abundances of
H and He to calculate $n^{\phantom1}_{\rm He}/n^{\phantom1}_{\rm H}$.
This is an approximation, since stellar evolution will increase
the helium content relative to hydrogen. 
We get $\mu^{\phantom1}_{\rm pr,H} = 1.3990$. 
We can then rewrite equation~(\ref{A4}) as
\begin{equation}
\label{A6}
Z_G={\mu^{\phantom1}_{Z,{\rm O}}\over\displaystyle
(\nH/\nO)\mu^{\phantom1}_{\rm pr,H}
+\mu^{\phantom1}_{Z,{\rm O}}}\,.
\end{equation}

Using the ratio $\nH/\nO$ 
given by equation~(\ref{eq-metal}), we can then calculate $Z_G$ for a
given galaxy stellar mass $M_\star$. 

%

\end{document}